\documentclass[journal]{IEEEtran}
\makeatletter

\usepackage{blindtext}
\usepackage[T1]{fontenc}
\usepackage{textcomp}
\usepackage{amsmath}
\usepackage{amssymb}
\usepackage{bm}
\usepackage{mdwmath}
\usepackage{cases}
\usepackage[short, c2]{optidef}
\usepackage{graphicx}
\graphicspath{{Figure/PDF/}{Biography/PDF/}}
\usepackage[dvipsnames]{xcolor}
\definecolor{myblue}{rgb}{0,0.4980,1} 
\definecolor{myred}{rgb}{0.8706,0.1608,0.0627} 
\usepackage[caption=false,font=footnotesize,subrefformat=parens]{subfig}
\usepackage[nosort,noadjust]{cite}
\usepackage{url}
\usepackage{tabularray}
\UseTblrLibrary{counter}
\usepackage[linesnumbered,ruled]{algorithm2e}
\usepackage{multirow}
\usepackage{comment}
\usepackage{hyperref}
\newcommand{\colorhypersetup}{\@ifpackageloaded{hyperref}{\hypersetup{%
	bookmarksopen=true,%
	bookmarksnumbered=true,%
	pdfpagemode={UseOutlines},
	pdfstartview={FitH},%
	colorlinks=true,%
	linkcolor={myred},%
	citecolor={orange}
}}{\empty}}
\newcommand{\blackhypersetup}{\@ifpackageloaded{hyperref}{\hypersetup{%
	bookmarksopen=true,%
	bookmarksnumbered=true,%
	pdfpagemode={UseOutlines},
	pdfstartview={FitH},%
	colorlinks=true,%
	allcolors={black}
}}{\empty}}
 \blackhypersetup

\usepackage{mfirstuc-english}
\MFUhyphentrue
\usepackage[version3,description,IEEEtran]{eacro}
\acsetup{%
    pages/display=all,%
    pages/seq/use=false,%
    pages/name=true,%
    pages/fill={\quad},%
     make-links=false%
}
\DeclareAcronym{isac}{
	short = ISAC,
	long = integrated sensing and communication}
 \DeclareAcronym{dt}{
	short = DA,
	long = digital agent}
  \DeclareAcronym{qoe}{
	short = QoE,
	long = quality of experience}
   \DeclareAcronym{qos}{
	short = QoS,
	long = quality of service}
    \DeclareAcronym{lru}{
	short = LRU,
	long = least recently used}
 \DeclareAcronym{los}{
	short = LoS,
	long = line of sight}
 \DeclareAcronym{aoa}{
	short = AoA,
	long = angle of arrival}
\DeclareAcronym{crb}{
	short = CRB,
	long = Cramér-Rao bound}
    \DeclareAcronym{mos}{
	short = MOS,
	long = mean opinion score}
\DeclareAcronym{arima}{
	short = ARIMA,
	long = autoregressive integrated moving average}
 \DeclareAcronym{mse}{
	short = MSE,
	long = mean square score}   
 \DeclareAcronym{drl}{
	short = DRL,
	long = deep reinforcement learning}  
\DeclareAcronym{sca}{
	short = SCA,
	long = successive convex optimization} 
\DeclareAcronym{bdqn}{
	short = BDQN,
	long = branch dueling Q network} 
 \DeclareAcronym{td3}{
	short = TD3,
	long = twin delayed deep deterministic policy gradient} 
     \DeclareAcronym{6g}{
	short = 6G,
	long = six-generation} 
     \DeclareAcronym{xr}{
	short = XR,
	long = extended reality} 
         \DeclareAcronym{csi}{
	short = CSI,
	long = channel state information} 
     \DeclareAcronym{snr}{
	short = SNR,
	long = signal-to-noise ratio} 
      \DeclareAcronym{vec}{
	short = VEC,
	long = vehicular edge computing} 
    \DeclareAcronym{ofdma}{
	short = OFDMA,
	long = orthogonal frequency division multiple access}
    \DeclareAcronym{tdma}{
	short = TDMA,
	long = time division multiple access}
    \DeclareAcronym{ssim}{
	short = SSIM,
	long = structural similarity index measure}  
\DeclareAcronym{td}{
	short = TD,
	long = temporal difference}

\newtheorem{lemma}{\textbf{Lemma}}

\newcommand{\upperroman}[1]{\uppercase\expandafter{\romannumeral#1}}

\newcommand{\myunit}[1]{%
	\ifmmode
		\mathrm{#1}
	\else
		$ \mathrm{#1} $
	\fi}
\newcommand{\murm}{%
	\ifmmode
		\text{\textmu}
	\else
		\textmu
	\fi}

\newcommand{\MYnewpage}{%
	\ifCLASSOPTIONonecolumn
		\ifCLASSOPTIONjournal
			\typeout{The onecolumn journal mode.}
			\newpage
		\fi
	\fi}

\newlength{\mysinglefigwidth}
\newlength{\mymultifigwidth}
\ifCLASSOPTIONonecolumn
	\AtBeginDocument{\setlength{\mysinglefigwidth}{0.7\linewidth}}
\else
	\AtBeginDocument{\setlength{\mysinglefigwidth}{\linewidth}}
\fi
\ifCLASSOPTIONonecolumn
	\AtBeginDocument{\setlength{\mymultifigwidth}{0.5\linewidth}}
\else
	\AtBeginDocument{\setlength{\mymultifigwidth}{\linewidth}}
\fi

\makeatother
\begin{document}
\ifCLASSOPTIONonecolumn
    \typeout{The onecolumn mode.}
    \title{Experience-Centric Resource Management in ISAC Networks: A Digital Agent-Assisted Approach}
        \author{Xinyu Huang, Yixiao Zhang, Yingying Pei, Jianzhe Xue, Weihua Zhuang,~\IEEEmembership{Fellow, IEEE}, and Xuemin (Sherman) Shen,~\IEEEmembership{Fellow, IEEE}
    \thanks{Xinyu Huang, Yixiao Zhang, Yingying Pei, Weihua Zhuang, and Xuemin (Sherman) Shen are with the Department of Electrical and Computer Engineering, University of Waterloo, Waterloo, ON N2L 3G1, Canada (E-mail: \{x357huan, y3549zha, y32pei, wzhuang, sshen\}@uwaterloo.ca).
    }
    \thanks{Jianzhe Xue is with the Department of Electrical and Computer Engineering, University of Waterloo, Waterloo, ON N2L 3G1, Canada, and with the School of Electronic Science and Engineering, Nanjing University, Nanjing, China. (Email: j59xue@uwaterloo.ca).}
    }

    
\else
    \typeout{The twocolumn mode.}
    \title{Experience-Centric Resource Management in ISAC Networks: A Digital Agent-Assisted Approach}
    \author{Xinyu Huang,~\IEEEmembership{Graduate Student Member, IEEE}, Yixiao Zhang,~\IEEEmembership{Graduate Student Member, IEEE},\\ Yingying Pei,~\IEEEmembership{Graduate Student Member, IEEE}, Jianzhe Xue,~\IEEEmembership{Graduate Student Member, IEEE},\\ and Xuemin (Sherman) Shen,~\IEEEmembership{Fellow, IEEE}
    \thanks{Xinyu Huang, Yixiao Zhang, Yingying Pei, and Xuemin (Sherman) Shen are with the Department of Electrical and Computer Engineering, University of Waterloo, Waterloo, ON N2L 3G1, Canada (E-mail: \{x357huan, y3549zha, y32pei, sshen\}@uwaterloo.ca).
    }
    \thanks{Jianzhe Xue is with the Department of Electrical and Computer Engineering, University of Waterloo, Waterloo, ON N2L 3G1, Canada, and with the School of Electronic Science and Engineering, Nanjing University, Nanjing, 210023, China. (Email: j59xue@uwaterloo.ca, jianzhexue@smail.nju.edu.cn).}
    }
    
    
\fi

\ifCLASSOPTIONonecolumn
	\typeout{The onecolumn mode.}
\else
	\typeout{The twocolumn mode.}
\fi

\maketitle

\ifCLASSOPTIONonecolumn
	\typeout{The onecolumn mode.}
	\vspace*{-50pt}
\else
	\typeout{The twocolumn mode.}
\fi
\begin{abstract}

In this paper, we propose a digital agent (DA)-assisted resource management scheme for enhanced user quality of experience (QoE) in integrated sensing and communication (ISAC) networks. Particularly, user QoE is a comprehensive metric that integrates quality of service (QoS), user behavioral dynamics, and environmental complexity. The novel DA module includes a user status prediction model, a QoS factor selection model, and a QoE fitting model, which analyzes historical user status data to construct and update user-specific QoE models. Users are clustered into different groups based on their QoE models. A Cramér-Rao bound (CRB) model is utilized to quantify the impact of allocated communication resources on sensing accuracy. A joint optimization problem of communication and computing resource management is formulated to maximize long-term user QoE while satisfying CRB and resource constraints. A two-layer data-model-driven algorithm is developed to solve the formulated problem, where the top layer utilizes an advanced deep reinforcement learning algorithm to make group-level decisions, and the bottom layer uses convex optimization techniques to make user-level decisions. Simulation results based on a real-world dataset demonstrate that the proposed DA-assisted resource management scheme outperforms benchmark schemes in terms of user QoE.
\end{abstract}

\ifCLASSOPTIONonecolumn
	\typeout{The onecolumn mode.}
	\vspace*{-10pt}
\else
	\typeout{The twocolumn mode.}
\fi
\begin{IEEEkeywords}
Digital agent, Cramér-Rao bound (CRB), quality of experience (QoE), resource management, integrated sensing and communication (ISAC).
\end{IEEEkeywords}

\IEEEpeerreviewmaketitle

\MYnewpage


\section{Introduction}
\label{sec:Introduction}
\Ac{isac} is expected to become a cornerstone of \ac{6g} networks, which can seamlessly integrate communication and sensing functionalities within a single communication network infrastructure~\cite{9585321,Xu2023}. The \ac{isac} systems usually leverage waveform design, joint signal processing, and resource sharing to optimize the performance of dual functionalities~\cite{9737357}. By reusing radio spectrum, transmit power, and hardware resources, the \ac{isac} systems aim at enhanced spectral efficiency and reduced system costs. The \ac{isac} service is pivotal for scenarios such as autonomous driving where vehicles rely on seamless communications and high-precision environmental sensing for real-time vehicle control decisions~\cite{10554983}, and mobile video streaming where the integrated system ensures accurate spatial awareness for immersive user experience~\cite{10216343}.

Within \ac{isac} networks, achieving a good trade-off between communication and sensing functionalities necessitates innovative resource management strategies due to their competition for limited resources, such as radio spectrum, transmit power, and computing resources~\cite{10158711}. Service-centric approaches often fail to address the diverse and dynamic requirements of \ac{isac} applications, where user priorities of communication and sensing can vary significantly~\cite{9982429}. Experience-centric resource management prioritizes user-level experience, i.e., \ac{qoe}, rather than \ac{qos} such as in terms of throughput or latency. To achieve the experience-centric resource management, advanced data processing and scheduling algorithms have been developed. For instance, data-driven predictive models are utilized to forecast user resource demands based on real-time user status and behavior patterns~\cite{10363344}. Adaptive priority-based strategies dynamically distribute resources by constantly reevaluating each user requirement and status to ensure an effective balance between communication and sensing performance~\cite{dong2022sensing}. However, achieving experience-centric resource management requires efficient and accurate user data management. 

To efficiently and accurately manage user data, the concept of \ac{dt} emerges as a key enabler. Building upon the foundation of digital twins, \ac{dt} represents an evolutionary step that integrates advanced data management functionalities tailored for dynamic and experience-centric resource management~\cite{shen2024revolutionizing}. \acp{dt} are expected to not only emulate user status in real-time but also extract fine-grained user characteristics, such as behavior patterns, which can enable precise modeling of user requirements in a complex and dynamic environment. For example, \acp{dt} can leverage multi-modal data sources, such as communication logs, sensor feedback, and application-level interactions, to generate comprehensive user profiles consisting of static attributes, e.g., user preferences, and dynamic states, e.g., mobility trajectories and network traffic variations. Furthermore, \acp{dt} possess advanced decision-making capabilities through machine learning and optimization techniques, which can dynamically optimize resource allocation based on user requirements and predicted resource demand fluctuations~\cite{10363344}.

Achieving experience-centric resource management in ISAC networks faces significant challenges. Firstly, user-specific \ac{qoe} models not only vary across users but also evolve over time due to dynamic user behaviors and changing environmental conditions, leading to difficulties in model establishment and update. Secondly, since resource optimization variables are typically coupled with complex and interdependent relationships, how to construct an efficient communication and sensing optimization framework to enhance user \ac{qoe} is an open research problem. Thirdly, large user scale and numerous optimization variables that need to be jointly optimized result in slow convergence of traditional optimization methods to an optimal or near-optimal solution. On the other hand, solely data-driven methods are highly susceptible to data drift.

In this paper, we propose a novel \ac{dt}-assisted resource management scheme in \ac{isac} networks, which can efficiently manage user-specific QoE models and make tailored resource scheduling strategies to enhance long-term user \ac{qoe}. The main contributions are summarized as follows:
\begin{itemize}
    \item 
    Firstly, we propose a \ac{dt} module to manage user-specific \ac{qoe} models. Within the DA module, a distance correlation coefficient (DCC) model is utilized to select the key \ac{qos} factors for each user QoE. To capture the dynamics of user behaviors and environmental complexity, a lightweight prediction model is leveraged. Based on the selected \ac{qos} factors, along with the predicted user behavioral dynamics and environmental complexity, a personalized \ac{qoe} model is constructed by data fitting and continuously refined through user feedback. Users with the same QoE model structure are clustered into a user group.
    \item 
    Secondly, we utilize \ac{crb} to model the impact of allocated communication resources on sensing accuracy. Particularly, each \ac{isac} user has three \ac{crb} constraints in terms of distance, velocity, and azimuth angle to achieve sensing accuracy. A joint optimization problem of communication and computing resource management to maximize long-term user \ac{qoe} is formulated by incorporating both \ac{crb} and resource constraints.
    \item 
    Thirdly, we propose a two-layer data-model-driven algorithm to solve the joint communication and computing resource optimization problem. In the top layer, an advanced \ac{drl} algorithm is utilized to make group-level network resource allocation decisions. In the bottom layer, the user-level resource allocation problem in each group is transformed into a convex problem solved by an optimization solver. Experiments based on a real-world dataset demonstrate the effectiveness of the proposed algorithm under different network resources and sensing requirements.
\end{itemize}

The remainder of this paper is organized as follows. Related works are reviewed in Section~\ref{related}. The system model, comprising a sensing model and a \ac{dt}-assisted \ac{qoe} model, is established, and a joint optimization problem is formulated in Section~\ref{system_model}. A two-layer data-model-driven algorithm is presented in Section~\ref{alg}, followed by simulation results in Section~\ref{sim}. Finally, Section~\ref{sec:Conclusion} concludes this study.

\section{Related Works}\label{related}
Resource management in \ac{isac} networks has been extensively studied in recent years. In the early days, communication resource management for \ac{isac} mainly focused on non-overlapping approaches, i.e., \ac{ofdma} and \ac{tdma}. In the \ac{ofdma} approaches, the frequency domain is partitioned into multiple subcarriers, with each subcarrier allocated to either sensing or communication~\cite{liu2017adaptive, liu2018mu, 7962141}. In the \ac{tdma} approaches, a time slot is divided into multiple mini-slots, with all subcarriers in each mini-slot allocated to either sensing or communication~\cite{xie2023optimal, xu2024dynamic, zhao2022integrated}. Generally, objectives for sensing mainly include target detection probability, sensing coverage, and sensing continuity~\cite{dong2022sensing,gan2024coverage}, while objectives for communications mainly include \ac{qos} such as in terms of delay and throughput~\cite{liu2024senscap}, and \ac{qoe} such as in terms of content quality and rebuffer time~\cite{10.1145}. Recently, a unified ISAC waveform design has attracted much attention. A single signal waveform is expected to simultaneously facilitate the sensing and communication functionalities to achieve high spectrum efficiency~\cite{dong2022sensing,10637248,an2023fundamental}. Intuitively, more spectrum resources can achieve better sensing and communication performance. However, spectrum resources are usually limited, and how to efficiently allocate them to achieve satisfactory performance is essential. To address this issue, researchers proposed various \ac{crb} models to depict the mathematical relationship between allocated communication resources and sensing performance~\cite{dong2022sensing, hua2023mimo}. To guarantee sensing accuracy, sensing \ac{crb} usually needs to remain below a prescribed threshold. Additionally, some existing works integrate communication resource management with computing or caching resource management to further enhance QoS or QoE~\cite{wen2024integrated, he2023integrated}. Since the integration introduces additional optimization variables in a complex network environment, advanced \ac{drl} algorithms have been adopted to make resource management decisions by learning the network state transition probabilities. 

Recently, \acp{dt} have attracted widespread attention on improving resource management performance.  A \ac{dt} can be deemed an intelligent proxy of a physical entity on the network side to perform efficient system behavior emulation and data analytics for customized resource management~\cite{shen2024revolutionizing}. The emulated behaviors mainly consist of user behaviors, including mobility trajectory and application interaction, and network operation behaviors, including topology change and network configuration~\cite{3gpp28915}. The behaviors can be emulated by prediction-based algorithms in network simulators, e.g., NS-3 and OPENT. Additionally, \acp{dt} can abstract useful network characteristics, such as spatiotemporal traffic variations~\cite{10228190}, swipe probability distributions~\cite{10363344}, and collaborative sensing demands~\cite{10643661}, from emulated behaviors and realistic environments. The information in \acp{dt} should help eliminate certain assumed prior knowledge in resource management and adapt to network dynamics for customized resource management.

In summary, most existing works on resource management in ISAC networks focus on performance enhancement based on universal QoE models and solely data-driven or model-based resource scheduling algorithms. Different from the existing studies, our \ac{dt}-assisted resource management scheme constructs and updates user-specific QoE models that integrate the impact of behavioral dynamics and environmental complexity on selected QoS factors. A hybrid data-model-driven resource scheduling algorithm decouples the joint optimization problem for dimension reduction and utilizes convex optimization to improve DRL robustness.

\section{System Model and Problem Formulation}\label{system_model}
In this section, a sensing model and a DA-assisted QoE model are established in an ISAC scenario. With the models, an experience-centric optimization problem is formulated.

\subsection{ISAC Scenario}
\begin{figure}[!t]
    \centering
    \includegraphics[width=0.9\mysinglefigwidth]{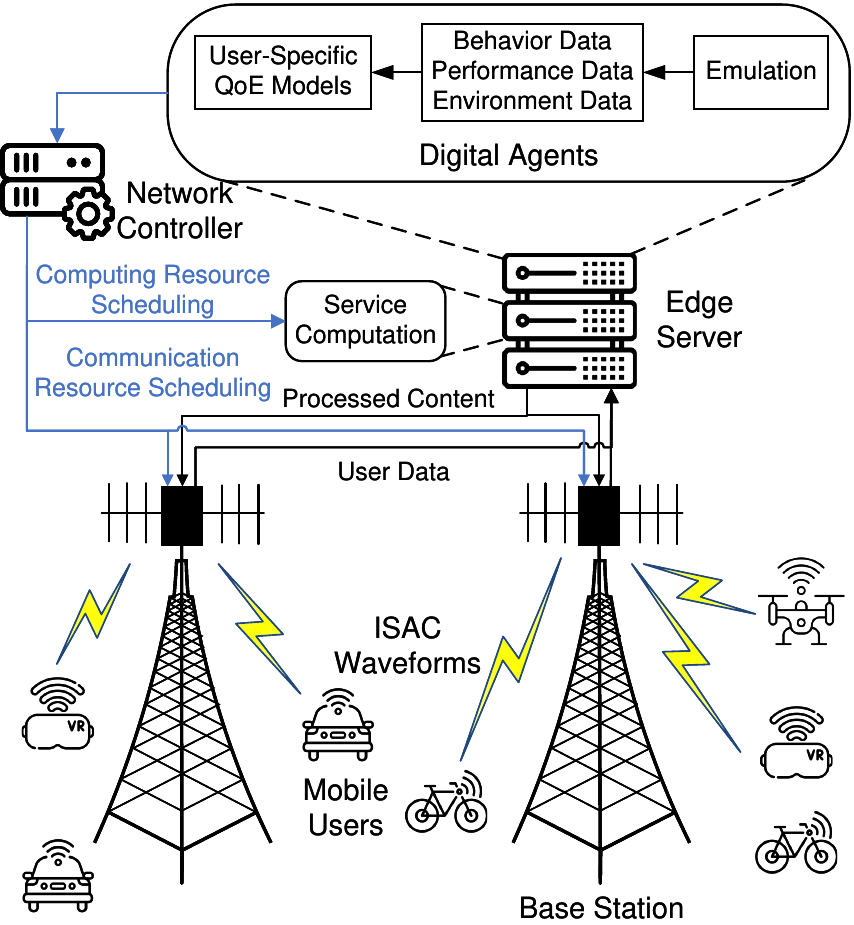}
    \caption{An illustration of DA-assisted resource management in the ISAC scenario.}
    \label{system}
\end{figure}

As shown in Fig.~\ref{system}, we consider a \ac{dt}-assisted resource management framework for the \ac{isac} scenario. The communication network mainly consists of two base stations, one edge server, multiple mobile users and their \acp{dt}, and one network controller. The base stations transmit unified ISAC waveforms to mobile users and analyze the reflected signal echoes to sense mobile user locations and movement speeds. The ISAC waveforms can support simultaneous data transmission and target sensing. Radio spectrum is partitioned to support the transmission of ISAC waveforms\footnote{In this work, we consider downlink transmission optimization for ISAC waveforms.}. User data, including behavior data, performance data, and environment data, are collected periodically for \ac{dt} update. Each mobile user has a unique \ac{dt} in the edge server, which predicts user status and abstracts a user-specific QoE model. The QoE model mainly consists of user behavioral dynamics, environmental complexity, and QoS such as in terms of service latency and content quality.
The mobile users are indexed by $k \in \{1,\cdots,K\}$,  with each user having a random but distinct velocity and trajectory. The edge server with substantial caching\footnote{In this work, we consider a fixed caching strategy, i.e., \ac{lru}, to update DA data and models for data freshness and model accuracy, and focus on communication and computing resource management.} and computing resources hosts \acp{dt} and processes user requests, such as video rendering and data analysis. The network operates in a time-slotted manner, in which the time slots are indexed by $n \in \{1,\cdots,N\}$. Assume that the variables are constant in each time slot and the carrier number exceeds the number of mobile users to avoid transmission interference. The network controller is responsible for communication and computing resource management based on real-time user status and their QoE models.

The main notations and their definitions are listed in Table~\ref{notation}.

\begin{table}[!t]
\centering
\caption{Main Notations and Definitions}
\label{notation}
\begin{tblr}{
    width = \linewidth,
    colspec = {X[0.32,c,c]X[2.7,c,c]},
    row{1} = {mode=text},
    row{1} = {font=\bfseries},
    hlines,
    hline{2} = {1}{-}{},
    hline{2} = {2}{-}{},
    vline{2}
}
    Notation & Definition  \\
    $\mathbf{r}_n(t)$ & The reflected signal echoes at time instance $t$ in time slot $n$ \\
    $B^\text{NN}$ & Null-to-null beam width of a receive antenna \\
    $B_{n,k}^\text{RMS}$ & User $k$'s effective bandwidth in time slot $n$ \\
    $H_{n,k}$ & User $k$'s behavioral dynamics in time slot $n$ (Sec. \ref{def}) \\
    $E_{n,k}$ & User $k$'s environmental complexity in time slot $n$ (Sec.\hspace{0.3mm}\ref{def}) \\
    $\mathcal{E}_{n,k}$ & User $k$'s QoE in time slot $n$ (Sec. \ref{def})\\
    $\mathcal{S}_{n,k}$ & User $k$'s QoS in time slot $n$ (Sec. \ref{def}) \\
    $\Upsilon_{n,k}$ & User $k$'s sensing CRB in time slot $n$ \\
    $B_{n,k}$ & The allocated bandwidth to user $k$ in time slot $n$ \\
    $P_{n,k}$ & The allocated transmit power to user $k$ in time slot $n$ \\
    $C_{n,k}$ & The allocated computing resources to user $k$ in time slot $n$ 
\end{tblr}
\end{table}

\subsection{Sensing Model}{
Each base station is equipped with $N_\mathtt{t}$ transmit antennas and $N_\mathtt{r}$ receive antennas to perform sensing and communication simultaneously. The baseband signals transmitted by the base stations are denoted by ${{\widehat{\mathbf{s}}}_{n}}\left( t \right)={{\left[ {{\widehat{s}}_{1,n}}\left( t \right),\ldots ,{{\widehat{s}}_{K,n}}\left( t \right) \right]}^{\text{T}}}\in {{\mathbb{C}}^{K\times 1}}$, where $t \in \mathbb{N}^+$ is the time instance index and one time slot includes multiple time instances. Superscript $\text{T}$ is matrix transpose. The transmitted signals need to be calibrated on amplitudes and phases across antennas through beamforming technology to enhance transmission efficiency, which can be expressed as ${\mathbf{s}_{n}}(t)={\mathbf{W}_{n}}{{\widehat{\mathbf{s}}}_{n}}(t)\in {{\mathbb{C}}^{{N_\mathtt{t}}\times 1}}$. Here, ${\mathbf{W}_{n}}=\left[ {\mathbf{w}_{1,n}},\ldots ,{\mathbf{w}_{K,n}} \right]\in {{\mathbb{C}}^{{N_\mathtt{t}}\times K}}$ is the downlink beamforming matrix in time slot~$n$. The reflected signal echoes from mobile users at time instance $t$ in time slot $n$ can be expressed as~\cite{liu2020radar}
\begin{align}\label{echo}
    {\mathbf{r}_{n}}(t)&=G\sum\limits_{k=1}^{K}{{{\beta }_{n,k}}{{e}^{j2\pi {{\mu }_{n,k}}t}}\mathbf{b}({{\theta }_{n,k}}){\mathbf{a}^{\text{H}}}({{\theta }_{n,k}}){\mathbf{s}_{n}}(t-{{v}_{n,k}})}\notag\\&+\mathbf{z}_n(t),
\end{align}
where $G=\sqrt{N_\mathtt{t} N_\mathtt{r}}$ is the total antenna array gain and $\beta_{n,k}$ represents the reflection coefficient that is related to propagation loss. In Eq.~\eqref{echo}, $v_{n,k}$ is the propagation delay, $\mu_{n,k}$ denotes the Doppler frequency, $\theta_{n,k}$ is the azimuth angle between mobile user $k$ and the base station, and $\mathbf{z}_{n}(t)\in{\mathbb{C}}^{{N_\mathtt{r}}\times 1}$ represents the circularly symmetric complex Gaussian noise vector. In Eq.~\eqref{echo}, vectors $\mathbf{a}$ and $\mathbf{b}$ represent transmit and receive steering vectors, respectively. Superscript $\text{H}$ is Hermitian transpose.

To estimate the sensing accuracy of user $k$ on distance $c_{n,k}$, velocity $v_{n,k}$, and azimuth angle $\theta_{n,k}$ in time slot $n$, we select \ac{crb} that is a lower bound on the variance of an unbiased estimator. The estimation \acp{crb} for sensing mobile user $k$ in time slot $n$ can be expressed as~\cite{dong2022sensing}
\begin{equation}\label{crb}
    \left\{ \begin{aligned}
  & {\Upsilon}_{n,k}^{(1)}=\frac{{{\lambda }_{1}}}{{{P}_{n,k}}{{\left| {{\varsigma }_{n,k}} \right|}^{2}}{{\left| B_{{n,k}}^{\text{RMS}} \right|}^{2}}}, \\ 
 & \Upsilon_{n,k}^{(2)}=\frac{{{\lambda }_{2}}}{{{P}_{n,k}}{{\left| {{\varsigma}_{n,k}} \right|}^{2}}{{\left| T_{{n,k}}^{\text{Eff}} \right|}^{2}}}, \\ 
 & \Upsilon_{n,k}^{(3)}=\frac{{{\lambda }_{3}}}{{{P}_{n,k}}{{\left| {{\varsigma }_{n,k}} \right|}^{2}}/{{B}^{\text{NN}}}}, \\ 
\end{aligned} \right.
\end{equation}
where $P_{n,k}$ and $\varsigma_{n,k}$ represent the transmit power and downlink sensing channel gain for mobile user $k$ in time slot $n$, respectively. In Eq.~\eqref{crb}, $\Upsilon_{n,k}^{(1)}, \Upsilon_{n,k}^{(2)}, \Upsilon_{n,k}^{(3)}$ represent the sensing performance of distance, velocity, and azimuth angle for the mobile user. A lower CRB value corresponds to a higher sensing accuracy.
Parameters $\lambda_1,\lambda_2,\lambda_3$ represent different sensing requirements for the distance, velocity and azimuth angle, respectively; $B^\text{NN}$ and $B_{n,k}^\text{RMS}$ represent a null-to-null beam width and an effective bandwidth for mobile user $k$, respectively. Effective bandwidth $B_{n,k}^\text{RMS}$ can be analyzed on the integral of a frequency-domain signal over its bandwidth~\cite{yan2015simultaneous}, given by 
\begin{equation}
    {{\left| B_{n,k}^{\text{RMS}} \right|}^{2}}=\frac{\int_{B_{n,k}}{{{f}^{2}}{{\left| {S}(f_{n,k}) \right|}^{2}}df}}{\int_{B_{n,k}}{{{\left| {S}(f_{n,k}) \right|}^{2}}df}},
\end{equation}
where $f_{n,k}$ is the carrier frequency for mobile user $k$ and $S(f_{n,k})$ is the Fourier transform of time-domain waveform ${{\widehat{\mathbf{s}}}_{n,k}}\left( \tilde{t} \right)$. Note that $\tilde{t}$ is the continuous form of $t$, and $S(f_{n,k})$ can be expressed as ${S}(f_{n,k})=\frac{1}{\pi {{f}_{n,k}}}\sin (\pi {{f}_{n,k}}{{T}^{\text{P}}})$~\cite{dong2022sensing}, where $T^\text{P}$ is the pulse width. Therefore, the effective bandwidth $B_{n,k}^\text{RMS}$ can be estimated as~\cite{dong2022sensing}
\begin{equation}
    {{\left| B_{n,k}^{\text{RMS}} \right|}^{2}}\approx B_{n,k}/\left( 2{{\pi }^{2}}{{T}^{\text{P}}} \right).
\end{equation}

In Eq.~\eqref{crb}, $T_k^\text{Eff}$ represents the effective time duration for mobile user $k$, which is used to quantify the temporal spread of the transmit signal's energy~\cite{yan2015simultaneous}. It can be expressed as
\begin{equation}
    {{\left| T_{n,k}^{\text{Eff}} \right|}^{2}}=\frac{\int_n{{\tilde{t}^{2}}{{\left| {{S}_{n,k}}(\tilde{t}) \right|}^{2}}d\tilde{t}}}{\int_n{{{\left| {{S}_{n,k}}(\tilde{t}) \right|}^{2}}d\tilde{t}}},
\end{equation}
where ${{S}_{n,k}}(\tilde{t})$ is the normalized complex envelope of the transmit signal for mobile user $k$, given by $S_{n,k}(\tilde{t}) ={\hat{s}_{n,k}(\tilde{t})}/{\sqrt{\int_n \left|\hat{s}_{n,k}(\tilde{t})\right|^2 d\tilde{t}}}$. Since the allocation of dwell time (the process of assigning a specific time interval during sensing stages) is not considered in this work, the effective time duration $T_{n,k}^\text{Eff}$ is set as a system parameter.

}

\subsection{DA-Assisted QoE Model}
In this subsection, we discuss how to utilize \acp{dt} to construct \ac{qoe} models for individual users.

\subsubsection{DA module overview}
\begin{figure}[!t]
    \centering
    \includegraphics[width=\mysinglefigwidth]{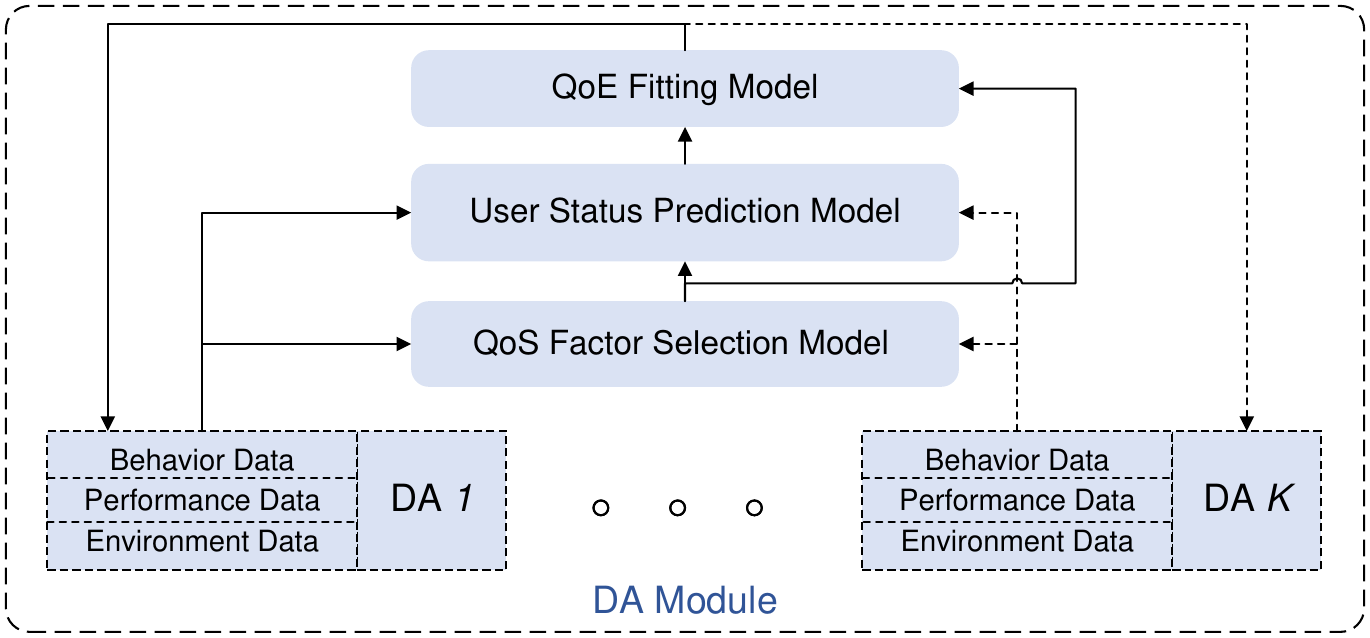}
    \caption{The operational mechanism in the DA module.}
    \label{DA-system}
\end{figure}
As shown in Fig.~\ref{DA-system}, a \ac{dt} module consists of multiple \acp{dt}, each of which stores the corresponding user behavior data, e.g., velocity and engagement time, performance data, e.g., service delay and content quality, and environment data, e.g., location and surrounding objects. The DA module is assumed to be a trusted data storage and processing module in communication networks. The DA data processing functionalities include a QoS factor selection model, a user status prediction model, and a QoE fitting model. In the QoS factor selection model, the input is service performance data from DAs, such as content quality and jitter, which are typically multi-dimensional and time-series. Directly using them to update a user QoE model can cause fitting distortion. Therefore, an efficient QoS factor selection model is essential, as it can justify which QoS factors can affect the current QoE model by using a distance correlation coefficient model. The output of QoS factor selection model is selected QoS factors for each user. In the user status prediction model, the input and output are historical DA data and predicted one, respectively. By using a lightweight prediction model, the future user status can be predicted with low resource consumption. In the QoE fitting model, the input includes the selected QoS factors and emulated contextual information from DA data. Considering the spatiotemporal dynamics in user behaviors and environment, it is important to incorporate their impact into the QoE model because users under different contexts usually have diversified service demands. With the determined QoS factors and emulated contextual information, a comprehensive QoE model can be established for each individual user. The DA data are used to trigger the QoE model update when the user experience is lower than a threshold.

\subsubsection{DA data collection}
DA data are classified into three categories, i.e., user behavior data, service performance data, and environment data, which are collected from physical networks and the user status prediction model in the DA module. DA data usually exhibit user variability and spatiotemporal discrepancies during data generation. Specifically, different users usually engage with various applications, e.g., social media, streaming, and online gaming, in varying ways, which leads to distinct behavioral patterns. Some users have long and continuous sessions, while others interact in short bursts. This affects the data generation rate and distribution. For example, in short video streaming services, some users may have frequent swipe behaviors while others may watch complete short videos in a sequential manner. This discrepancy can lead to different swipe-behavior uploads. The user status prediction model in the DA module can help generate DA data to guarantee that user status can always be aware on the network side, but it also brings some prediction bias due to data drift. More data needs to be collected from physical networks for the user status prediction model update through incremental learning algorithms when the average prediction accuracy is below a prescribed threshold, as specified in~\cite{huang2024adaptive}. With the help of the user status prediction model, DA data collection can be adaptive to user status variation patterns. When the user status variation pattern is relatively static, the user status prediction model can have high accuracy and DA data collection can fully rely on the user status prediction model. Otherwise, DA data need to be collected from physical networks to guarantee user status can be aware in real time and update the user status prediction model. Based on the preceding analysis, we propose an adaptive DA data collection frequency equation, given by
\begin{equation}
    {{l}_{k,i}}=\overline{l}_{i}{{e}^{-{{\nu}_{i}}{{A}_{k,i}}}},
\end{equation}
where $l_{k,i}$ denotes the DA data collection frequency for data attribute $i$ from user $k$. Parameter $\overline{l}_{i}$ is the standard data collection frequency for data attribute $i$, which can be retrieved from existing data collection standards, such as~\cite{3gpp37817,3gpp37816}. For example, in short video streaming services, user swipe behaviors are usually collected every 5 seconds; in map navigation services, user locations are usually collected every 1 second. Particularly, $e^{-\nu_{i} A_{k,i}}$ is a scaling factor to adaptively adjust data collection frequency, where $\nu_{i}$ and $A_{k,i}$ denote the attenuation rate of data collection frequency for data attribute~$i$ and average prediction accuracy for DA data attribute $i$ of user~$k$, respectively. With the increasing prediction accuracy, DA data collection frequency gradually decreases.

\subsubsection{QoS factor selection in the DA module}
As shown in Fig.~\ref{factor}, we present how to select QoS factors for each individual user in the DA module. Since the QoE reflects a user's overall satisfaction over large time durations, the user behavior data and service performance data, such as engagement time, content quality, and sensing accuracy, are collected in each large time duration, e.g, over $[T_0\sim T_1]$, to estimate the corresponding \ac{mos}. Content quality denotes the quality of transmitted data, which is usually measured when a complete data file is received in the user terminal. Therefore, it is usually a piecewise constant in the time domain. For example, in video streaming services, content quality can refer to \ac{ssim} of video segments or frames. The \ac{mos} usually ranges from $1\sim 5$, which has a prescribed mapping relationship with DA data~\cite{pezzulli2020estimation}. A higher \ac{mos} corresponds to a higher user experience.  In this work, we use \ac{mos} to measure user QoE. The content quality is equally split into several parts, with a higher content quality corresponding to a higher \ac{mos}. Then, the average QoS factors and QoE during each large time duration are used to obtain different matrices. The matrices are input to the distance correlation coefficient model to select QoS factors for user-specific QoE models. Let take the service delay matrix, denoted by $\mathbf{D}=[D_0,\cdots,D_{M-1}]$, and the QoE matrix, denoted by $\mathbf{Q}=[\mathcal{E}_0,\cdots,\mathcal{E}_{M-1}]$, as an example to introduce the operational procedure of the distance correlation coefficient model. In matrices $\bf{D}$ and $\bf{Q}$, $M$ is the number of selected large time durations, and $D_m$ and $\mathcal{E}_{m}$ are scalars. The specific procedure is as follows:
\begin{figure}[!t]
    \centering
    \includegraphics[width=\mysinglefigwidth]{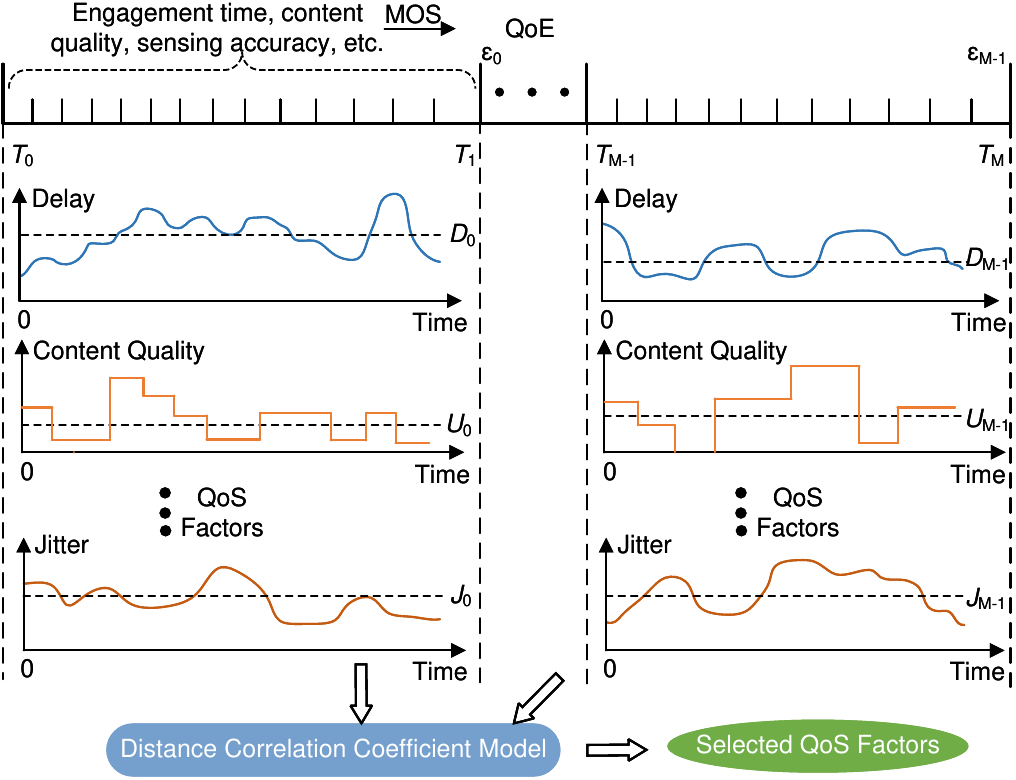}
    \caption{The QoS factor selection procedure in the DA module.}
    \label{factor}
\end{figure}

First, we need to calculate the pairwise Euclidean distance matrices, i.e., $\mathbf{X}$ for $\mathbf{D}$ and $\mathbf{Y}$ for $\mathbf{Q}$. The scalar element $X_{ij}$ in matrix $\mathbf{X}$ is defined as the absolute difference $|D_i - D_j|$ between scalar elements $D_i$ and $D_j$ in matrix $\mathbf{D}$, and similarly for $Y_{ij}$ in matrix~$\mathbf{Y}$. These matrices are then double-centered using the formula:
\begin{equation}
    X^*_{ij} = X_{ij} - \overline{X}_{i.} - \overline{X}_{.j} + \overline{X}_{..},
\end{equation}
where $\overline{X}_{i.}$ is the mean of the $i$-th row, $\overline{X}_{.j}$ is the mean of the $j$-th column, and $\overline{X}_{..}$ is the overall mean of matrix $\mathbf{X}$. The same principle is applied to matrix $\mathbf{Y}$.

The distance covariance $\mathcal{V}^2(\mathbf{D}, \mathbf{Q})$ is then calculated as
\begin{equation}
    \mathcal{V}^2(\mathbf{D}, \mathbf{Q}) = \frac{1}{M^2} \sum_{i=0}^{M-1} \sum_{j=0}^{M-1} X^*_{ij} Y^*_{ij}.
\end{equation}

Similarly, the distance variances of matrices $\mathbf{D}$ and $\mathbf{Q}$ are computed as
$\mathcal{V}^2(\mathbf{D}, \mathbf{D}) = \frac{1}{M^2} \sum_{i=0}^{M-1} \sum_{j=0}^{M-1} X^*_{ij} X^*_{ij}$ and $\mathcal{V}^2(\mathbf{Q}, \mathbf{Q}) = \frac{1}{M^2} \sum_{i=0}^{M-1} \sum_{j=0}^{M-1} Y^*_{ij} Y^*_{ij}$, respectively.

Finally, the DCC of matrices $\mathbf{D}$ and $\mathbf{Q}$ can be expressed as~\cite{dueck2017distance}
\begin{equation}
\mathcal{F}(\mathbf{D}, \mathbf{Q}) = \frac{\mathcal{V}(\mathbf{D}, \mathbf{Q})}{\sqrt{\mathcal{V}(\mathbf{D}, \mathbf{D}) \times \mathcal{V}(\mathbf{Q}, \mathbf{Q})}},    
\end{equation}
where the coefficient $\mathcal{F}(\mathbf{D}, \mathbf{Q})$, ranging from $0$ to $1$, quantifies the association strength between service delay and QoE. A high coefficient indicates a high association strength.

Based on the distance correlation coefficient model, the DA module can select which QoS factors are related to the user-specific QoE models, thereby improving the efficiency of QoE model establishment.

\subsubsection{User status prediction in the DA module}
The QoE model fitting is performed at the end of each large time duration, which requires the time-average user status information. Therefore, we construct a large-timescale prediction model in the DA module, which can alleviate the impact of uncertain behaviors and environment in the small timescale. To reduce computation consumption, we select a lightweight \ac{arima} method to predict the future user status. Take environmental complexity sequence $\hat{\mathtt{{E}}}= [E_1, \dotsc, E_M]$ as an example, the detailed steps of predicting $E_{M+1}$ are as follows:

First, we need to check the stationarity of the time series. If series \( \hat{\mathtt{{E}}} \) is non-stationary, we can transform it into a stationary state by differencing it $w$ times, and obtain the differenced series, given by
\begin{equation}
    y_m = \Delta^w E_m = (1 - L)^w E_m,
\end{equation}
where $L$ is the lag operator, and \( \Delta \) represents the differencing operation.

Then, we need to determine orders \( p \) and \( q \) of the ARIMA model. By analyzing the autocorrelation function (ACF) and partial ACF of differenced series \( y_m \), we can select appropriate autoregressive order \( p \) and moving average order \( q \) to establish the $\text{ARIMA} ( p, w, q )$ model. Specifically, When selecting $p$ and $q$, the ACF and partial ACF plots need to be observed, where a rapid cutoff in partial ACF suggests a suitable autoregressive order $p$, and a rapid cutoff in ACF indicates an appropriate moving average order $q$. By comparing candidate models using information criteria and residual analysis, the parameter combination can be selected to achieve a balance between prediction accuracy and computational efficiency.

Next, we need to estimate the model parameters. By using the maximum likelihood estimation method, we can estimate the autoregressive coefficient \( \phi_i \) and moving average coefficient \( \theta_j \). The ARIMA model is formulated as
\begin{equation}\label{phi}
    \Phi(L) y_m = \Theta(L) \varepsilon_m,
\end{equation}
where $\Phi(L) = 1 - \phi_1 L - \phi_2 L^2 - \dotsb - \phi_p L^p$ and $\Theta(L) = 1 + \theta_1 L + \theta_2 L^2 + \dotsb + \theta_q L^q$. In Eq.~\eqref{phi}, \( \varepsilon_m \) is the white noise error term.

After building the ARIMA model, differenced series \( \hat{y}_{M+1} \) can be predicted as
\begin{equation}
    \hat{y}_{M+1} = \sum_{i=1}^{p} \phi_i y_{M+1 - i} + \sum_{j=1}^{q} \theta_j \varepsilon_{M+1 - j}.
\end{equation}
Since future error terms are unknown, we typically assume \( \varepsilon_m = 0 \) for \( m > M \).

Finally, we can perform inverse differencing to recover the predicted value of the original series \( {E}_{M+1} \). Depending on the differencing order \( w \), the inverse differencing formula is expressed as

\begin{equation}
    \left\{
    \begin{aligned}
  &{E}_{M+1} = \hat{y}_{M+1} + E_M, w=1, \\
  &{E}_{M+1} = \hat{y}_{M+1} + 2E_M - E_{M-1}, w=2,
\end{aligned}
\right.
\end{equation}
where we only consider differencing order $w$ with values $1$ and $2$, because they are the most common cases encountered in practice, which can provide an accurate estimation.

Through these steps, an ARIMA model can be built at the DA module to make accurate and lightweight user status prediction.

\subsubsection{QoE model fitting \& update in the DA module} \label{def}

To accurately capture realistic user satisfaction with network services, a comprehensive QoE model is crucial. We consider a QoE model that integrates the impact of behavioral dynamics and environmental complexity on QoS. The behavioral dynamics is analyzed based on the variation of user behavior data, while the environmental complexity is estimated based on the number of surrounding buildings and obstacles in a specific location. When users remain relatively stationary in simpler environments, their QoS requirements can be relaxed. In contrast, in more dynamic or complex settings, stricter QoS requirements are necessary to maintain satisfactory user experience. This adaptability allows the QoE model to remain responsive to contextual factors, which can enable the network operator to dynamically adjust service quality to adapt to changing environmental conditions and user behaviors. Based on the preceding analysis, the comprehensive QoE model is constructed as
\begin{equation}\mathcal{E}_{n,k} = \mathcal{S}_{n,k} \times I(H_{n,k}, E_{n,k}),
\end{equation}
where $\mathcal{E}_{n,k}$ and $\mathcal{S}_{n,k}$ represent mobile user $k$'s QoE and QoS in time slot $n$, respectively. Function $I(\cdot)$ is an impact function and $H_{n,k}$ is mobile user $k$'s behavioral dynamics. Two kinds of QoS factors are considered in the DA-assisted QoS factor selection model, i.e., service delay and content quality, which have linear combination structure and nonlinear one~\cite{zheng2012qos, zou2012qos, kimura2016quve}, represented by $\mathtt{L}_1$ and $\mathtt{L}_2$. In each time slot, the QoS of mobile user $k$ can be expressed as
\begin{equation}\label{qos}
    \mathcal{S}_{n,k} =
     \left\{
    \begin{aligned}
    &\underbrace{\frac{\omega_{n,k}^{(1)}\xi_{k} F_{n,k}}{1+\xi_{k} F_{n,k}}}_{\text{content quality}} - \omega_{n,k}^{(2)}\underbrace{\left(\frac{\mu F_{n,k}}{C_{n,k}} + \frac{F_{n,k}}{R_{n,k}}\right)}_{\text{service latency}}, \mathtt{M}_{n,k}=\mathtt{L}_1, \\
    &\omega_{n,k}^{(3)}\frac{\xi_{k} F_{n,k}}{1+\xi_{k} F_{n,k}} / \left(\frac{\mu F_{n,k}}{C_{n,k}} + \frac{F_{n,k}}{R_{n,k}}\right), \mathtt{M}_{n,k}=\mathtt{L}_2,
    \end{aligned}
    \right.
\end{equation}
where the QoS model structure of user $k$ is denoted by $\mathtt{M}_{n,k}$, which can be determined by data fitting methods. The first component is content quality, which has a positive correlation with file size $F_{n,k}$~\cite{huang2020online, huang2024adaptive}. In Eq.~\eqref{qos}, $\xi_{k}$ is the weighting parameter in the content quality, whose value is typically set to $2$~\cite{huang2024adaptive}. The second component is service latency, where parameter $\mu$ is computing density and $R_{n,k}$ is the transmission capability of mobile user $k$ in time slot $n$. The transmission capability $R_{n,k}$ can be estimated by allocating bandwidth resources, given by
$R_{n,k}=B_{n,k} \log(1+P_{n,k} |h_{n,k}|^2/(B_{n,k} \sigma_r^2))$, where $h_{n,k}$ is the communication channel gain of mobile user $k$ and $\sigma_r^2$ is the noise power spectral density. To balance the content quality and service latency, we introduce three weighting factors $\omega_{n,k}^{(1)}$, $\omega_{n,k}^{(2)}$ and $\omega_{n,k}^{(3)}$, which can be fitted by the mapping between historical QoE and QoS. 

As analyzed before, impact function $I(\cdot)$ is regarded as an elastic factor influencing the QoS function, where high user dynamics and complex environment correspond to a strict QoS requirement, and vice versa. Since behavioral dynamics $H_{n,k}$ and environmental complexity $E_{n,k}$ have different units and ranges, typically larger than $0$, we need to normalize them, i.e., $\tilde{H}_{n,k} = H_{n,k} / H_{\text{max}}$ and $\tilde{E}_{n,k} = E_{n,k} / E_{\text{max}}$. Additionally, we assume that the impact function should have the following characteristics. When user dynamics and environmental complexity are $0$, the user QoS requirement remains unchanged. Conversely, as both increase, a stricter QoS requirement becomes apparent. Based on the preceding analysis, a parameterized Sigmoid function is utilized to characterize the impact function, given by
\begin{equation}
    I({H}_{n,k},{E}_{n,k}) = 1 - \frac{0.3}{1+e^{-5{(\tilde{H}_{n,k}+\tilde{E}_{n,k}}-1)}},
\end{equation}
where impact function $I(\cdot)$ ranges from $(0.7,1)$, and it is negatively correlated with ${H}_{n,k}$ and ${E}_{n,k}$. 

Since user behavior patterns are not always static, the dynamics require updating user QoE models, including model structures ($\mathtt{L}_1,\mathtt{L}_2$) and parameters ($\omega_{n,k}^{(1)}, \omega_{n,k}^{(2)}, \omega_{n,k}^{(3)},\forall k\in\mathcal{K}$). Specifically, the relevance of various factors is first evaluated through the DCC analysis to identify which factors are currently dominant. This ensures that the QoE model remains sensitive to the most pertinent aspects of user experience. Then, the model structures and parameters are updated using data fitting techniques to guarantee that the updated QoE has a low \ac{mse} compared with the one estimated by MOS.

\subsubsection{Data and model interaction within the DA module}
As shown in Fig.~\ref{DA-system}, there exists a closed loop between DA data and models. Specifically, DA data are input to the QoS factor selection model to identify which kinds of user status are essential for experience-centric network management. Based on the determined user status, the user status prediction model can sample them from historical DA data and predict their future values. The predicted user status is further input to the QoE fitting model to guarantee timeliness and improve accuracy. The fitted QoE models are finally stored in the corresponding DAs, which can be used for experience-centric network management.

\subsection{Experience-Centric Optimization Problem Formulation}

Since user behaviors and environmental complexity usually change over time, network resource allocation needs to adapt to the dynamics to guarantee communication and sensing performance. Given the limited communication and computing resources in the ISAC networks, our objective is to maximize long-term user QoE with guaranteed sensing performance. To achieve this goal, the network controller needs to accurately allocate bandwidths, transmit power, and computing resources to each user. Therefore, the formulated optimization problem can be expressed as
\begin{subequations}
\begin{align}
  & \underset{{{\left\{ {\textbf{B}_{n}},{\textbf{P}_{n}},{\textbf{C}_{n}} \right\}}_{n\in \mathcal{N}}}}{\mathop{\max }}\,\underset{N\to \infty }{\mathop{\lim }}\,\frac{1}{N}\sum\limits_{n=1}^{N}{\frac{1}{K}\sum\limits_{k=1}^{K}{\mathcal{E}_{n,k}}} \\ 
 \text{s.t.}\quad& \Upsilon_{n,k}^{(1)}\le {{\alpha }_{1}},\forall k\in \mathcal{K},n\in \mathcal{N}, \label{con1} \\ 
 & \Upsilon_{n,k}^{(2)}\le {{\alpha }_{2}},\forall k\in \mathcal{K},n\in \mathcal{N}, \label{con2} \\ 
 & \Upsilon_{n,k}^{(3)}\le {{\alpha }_{3}},\forall k\in \mathcal{K},n\in \mathcal{N}, \label{con3} \\ 
 & \sum\nolimits_{k=1}^{K}{{{B}_{n,k}}\le {{B}_\text{Total}}},\forall n\in \mathcal{N}, \label{con4}\\ 
 & \sum\nolimits_{k=1}^{K}{{{P}_{n,k}}\le {{P}_\text{Total}}},\forall n\in \mathcal{N}, \label{con5}\\ 
 & \sum\nolimits_{k=1}^{K}{{{C}_{n,k}}\le {{C}_\text{Total}}},\forall n\in \mathcal{N}, \label{con6}
\end{align}
\end{subequations}
where $\textbf{B}_n, \textbf{P}_n, \textbf{C}_n$ represent the bandwidth allocation matrix, transmit power allocation matrix, and computing resource allocation matrix for all mobile users, respectively. Parameters $\alpha_1, \alpha_2, \alpha_3$ denote the sensing CRB thresholds for distance, velocity, and azimuth angle, respectively. Constraints~\eqref{con4}-\eqref{con6} are the resource constraints that guarantee the allocated bandwidths, transmit power, and computing resources do not exceed the system capacity.

\section{Two-Layer Data-Model-Driven Solution}\label{alg}
In this section, we propose a two-layer data-model-driven algorithm to solve the formulated optimization problem.

\subsection{Overview of Proposed Solution}
Since users may have various QoE model structures with unique convexity or concavity properties, their combination can complicate the transformation of the objective function into a solely convex or concave form. Furthermore, due to the lack of future information, finding a global optimal resource allocation strategy via optimization methods is difficult. To approach a near-optimal solution, one way is to use \ac{drl} algorithms. However, directing inputting all user statuses into a \ac{drl} algorithm and making resource allocation decisions for each user can cause the curse of dimensionality~\cite{10302364}. 

\begin{figure}[!t]
    \centering
    \includegraphics[width=\mysinglefigwidth]{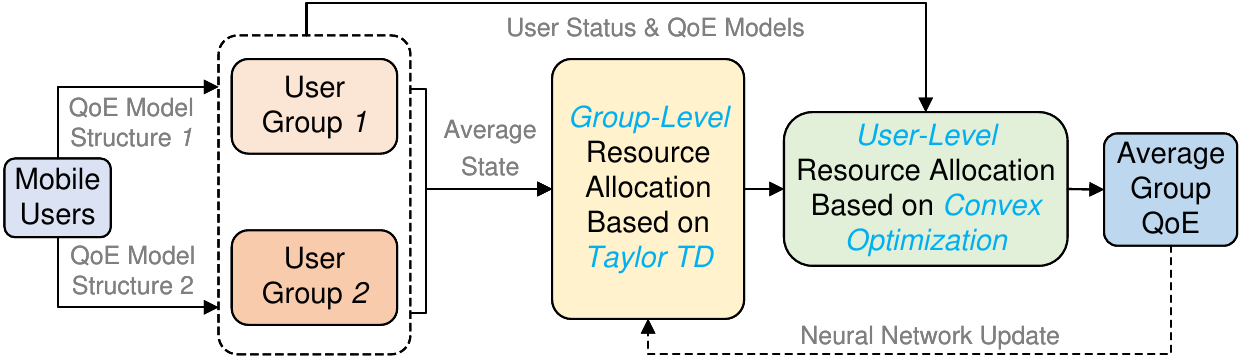}
    \caption{The proposed two-layer data-model-driven algorithm.}
    \label{hier}
\end{figure}

Therefore, we propose a two-layer data-model-driven algorithm, where the \ac{drl} algorithm is responsible for making group-level resource allocation decisions, and the convex optimization method handles user-level resource allocation problems, as shown in Fig.~\ref{hier}. The algorithm operates as follows.
First, mobile users are clustered into different groups based on their QoE model structures. Then, the average state of each user group is input to the Taylor \ac{td} algorithm~\cite{garibbo2024taylor} to make group-level bandwidth, transmit power, and computing resource allocation, which can reduce the dimensionality of state and action. Next, the user-level resource allocation problem of each group is transformed into a convex problem solved by convex optimization to achieve an optimal result. Finally, average group QoE is utilized to update the Taylor TD algorithm in the training stage. 

\subsection{DRL-Based Group-Level Resource Allocation}
In the proposed two-layer data-model-driven algorithm, we adopt the Taylor TD algorithm to determine group-level resource allocation. The Taylor TD algorithm enhances learning stability and accuracy through first-order Taylor corrections. Its ability to model subtle trade-offs and capture fine-grained changes can make it particularly suited for group-level resource allocation decisions, where dynamic and long-term optimization problems can be effectively solved. In the following, we will present the detailed state, action, reward, and neural network training mechanism in the Taylor TD algorithm.

\subsubsection{State}
The state consists of six kinds of elements, i.e., user group $i$'s average behavioral dynamics $\overline{H}_n^i$, average environmental complexity $\overline{E}_n^i$, average sensing channel gain $\overline{\varsigma}_n^i$, average communication channel gain $\overline{h}_n^i$, average file size $\overline{F}_n^i$, and average QoE model parameters $\overline{\bm{\omega}}_n^i$ at step $n$. The state can be expressed as
\begin{equation}
    \texttt{s}_n = \{\overline{H}_n^i, \overline{E}_n^i, \overline{\varsigma}_n^i, \overline{h}_n^i, \overline{F}_n^i, \overline{\bm{\omega}}_n^i\}_{i=1,2}.
\end{equation}

\subsubsection{Action}
The action includes bandwidth, transmit power, and computing resource allocation decisions for each group, which can be expressed as
\begin{equation}
    \texttt{a}_n = \{B_n^i, P_n^i, C_n^i\}_{i=1,2}.
\end{equation}
The action value ranges from $0$ to $1$. To guarantee that action can satisfy resource constraints, we need to normalize it. For example, the normalized bandwidth resource decision of group $i$ can be expressed as $\widehat{B}_n^i = B_\text{Total} \times B_n^i/\sum_i B_n^i$. A similar normalization operation also applies to transmit power allocation and computing resource allocation.

\subsubsection{Reward}
The reward selects average user QoE, i.e., $\texttt{r}_n=\frac{1}{K}\sum_k\mathcal{E}_{n,k}$, which can be obtained after user-level resource allocation.

\subsubsection{Neural network training mechanism}
Traditional TD learning algorithms update value function $\texttt{Q}(\texttt{s}, \texttt{a})$ based on the TD error, given by
\begin{equation}\label{td-1}
\delta = \texttt{r} + \gamma \max_{\texttt{a}'} \texttt{Q}(\texttt{s}', \texttt{a}') - \texttt{Q}(\texttt{s}, \texttt{a}),
\end{equation}
where $\delta$ represents the TD error, $\gamma$ is the discount factor, and $\texttt{Q}(\texttt{s}, \texttt{a})$ and $\texttt{Q}(\texttt{s}', \texttt{a}')$ are the current and next state-action value functions, respectively.

In Taylor TD learning, the update is refined with a first-order Taylor correction term, given by
\begin{equation}
\texttt{Q}(\texttt{s}, \texttt{a}) \approx \texttt{Q}(\texttt{s}, \texttt{a}) + \frac{\partial \texttt{Q}(\texttt{s}, \texttt{a})}{\partial \texttt{a}} \cdot \Delta \texttt{a},
\end{equation}
where $\frac{\partial \texttt{Q}(\texttt{s}, \texttt{a})}{\partial \texttt{a}}$ represents the gradient of $\texttt{Q}(\texttt{s}, \texttt{a})$ with respect to action $\texttt{a}$, and $\Delta \texttt{a}$ is the adjustment in the action. The updated value function is expressed as
\begin{equation}\label{td-2}
\texttt{Q}(\texttt{s}, \texttt{a}) \leftarrow \texttt{Q}(\texttt{s}, \texttt{a}) + \eta \left(\delta + \frac{\partial \texttt{Q}(\texttt{s}, \texttt{a})}{\partial \texttt{a}} \cdot \Delta \texttt{a}\right),
\end{equation}
where $\eta$ is the learning rate. The Taylor correction enables the algorithm to capture local trends in the value function, thereby improving update precision.

\subsection{Model-Based User-Level Resource Allocation}
At step $n$, based on the diversity of QoE model structures, the user-level optimization problem can be decomposed into two subproblems $\textbf{SP}_1$ and $\textbf{SP}_2$, as follows:
\begin{align}\label{sp1}
    \textbf{SP}_1:&\underset{{{\left\{ {\textbf{B}_{n}},{\textbf{P}_{n}},{\textbf{C}_{n}} \right\}}}}{\mathop{\max }}{\frac{1}{K_1}\sum\limits_{k=1}^{K_1}{\mathcal{E}_{n,k}^{(1)}}} \\
    &\text{s.t.} \quad \eqref{con1}- \eqref{con6}, \notag
\end{align}
\begin{align}\label{sp2}
    \textbf{SP}_2:&\underset{{{\left\{ {\textbf{B}_{n}},{\textbf{P}_{n}},{\textbf{C}_{n}} \right\}}}}{\mathop{\max }}{\frac{1}{K_2}\sum\limits_{k^{'}=1}^{K_2}{\mathcal{E}_{n,k^{'}}^{(2)}}} \\
    &\text{s.t.} \quad \eqref{con1}-\eqref{con6}, \notag
\end{align}
where $k\in\{1,\cdots,K_1\}$ and $k^{'}\in\{1,\cdots,K_2\}$ represent user indexes in user groups $1$ and $2$, respectively. In Eqs.~\eqref{sp1}\eqref{sp2}, $\mathcal{E}_{n,k}^{(1)}$ and $\mathcal{E}_{n,k'}^{(2)}$ are distinguished based on QoS model structures in Eq.~\eqref{qos}. The total group bandwidth, transmit power, and computing resource in $\textbf{SP}_1$ and $\textbf{SP}_2$ are obtained via the Taylor TD algorithm.

\begin{lemma}
\label{theo1}
Both subproblems $\textbf{SP}_1$ and $\textbf{SP}_2$ are non-convex optimization problems.
\end{lemma}
\begin{IEEEproof}
For subproblem $\textbf{SP}_1$, its objective function is expressed as
\begin{align}
\mathcal{E}_{n,k}^{(1)} = &I(H_{n,k}, E_{n,k}) \notag \times \\& \left( 1 - \frac{1}{1 + \xi_{k} F_{n,k}} - \omega_{n,k}^{(1)} \left( \frac{\mu F_{n,k}}{C_{n,k}} + \frac{F_{n,k}}{R_{n,k}} \right) \right).
\end{align}
Since $F_{n,k}$ is a constant term, the term $1 - \tfrac{1}{1 + \xi_{k} F_{n,k}}$ is also a constant. Let define $
\Gamma_{n,k} = I(H_{n,k}, E_{n,k}) \left( 1 - \frac{1}{1 + \xi_{k} F_{n,k}} \right)$, where $I(H_{n,k}, E_{n,k})$ is a constant term. Then, we can have $\mathcal{E}_{n,k}^{(1)} = \Gamma_{n,k} - I(H_{n,k}, E_{n,k}) \, \omega_{n,k}^{(1)} \left( \frac{\mu F_{n,k}}{C_{n,k}} + \frac{F_{n,k}}{R_{n,k}} \right).$ Next, let define $\widehat{\alpha}_{n,k} = I(H_{n,k}, E_{n,k}) \ \omega_{n,k}^{(1)} \mu F_{n,k}$ and $\widehat{\beta}_{n,k} = I(H_{n,k}, E_{n,k}) \, \omega_{n,k}^{(1)} F_{n,k}$. Therefore, QoE function in subproblem $\textbf{SP}_1$ can be rewritten as
\begin{equation}
    \mathcal{E}_{n,k}^{(1)} = \Gamma_{n,k} - \left( \frac{\widehat{\alpha}_{n,k}}{C_{n,k}} + \frac{\widehat{\beta}_{n,k}}{R_{n,k}} \right).
\end{equation}

Maximizing $\mathcal{E}_{n,k}^{(1)}$ w.r.t. $B_{n,k}$, $P_{n,k}$, and $C_{n,k}$ depends on minimizing $\tfrac{1}{C_{n,k}}$ and $\tfrac{1}{R_{n,k}}$.
\begin{itemize}
    \item Convexity of $\frac{1}{C_{n,k}}$: Function $f(C_{n,k}) = \frac{1}{C_{n,k}}$ is convex for $C_{n,k} > 0$, because $f''(C_{n,k}) = \frac{2}{C_{n,k}^3} > 0$. However, in the objective, we have $-\frac{\widehat{\alpha}_{n,k}}{C_{n,k}}$. Therefore, it is concave in terms of $C_{n,k}$.
    \item Convexity of $\frac{1}{R_{n,k}}$: Considering the transmission capability function is expressed as $R_{n,k} = B_{n,k} \log_2\left(1 + \frac{P_{n,k} |h_{n,k}|^2}{B_{n,k} \sigma_r^2}\right)$, the function involves $B_{n,k}$ and $P_{n,k}$ in a non-trivial way. Let define $a = \frac{|h_{n,k}|^2}{\sigma_r^2}$, we can have $R_{n,k} = B_{n,k} \log_2\left(1 + \frac{a P_{n,k}}{B_{n,k}}\right)$. The Hessian of $R_{n,k}$ w.r.t. $(B_{n,k}, P_{n,k})$ is neither positive semidefinite nor negative semidefinite. Hence, $R_{n,k}$ is neither convex nor concave. Consequently, $\frac{1}{R_{n,k}}$ is not convex. Thus, the term $\frac{\beta_{n,k}}{R_{n,k}}$ is non-convex in terms of $B_{n,k}$ and $P_{n,k}$.
\end{itemize}

Since the objective function of subproblem $\textbf{SP}_1$ includes a non-convex term $\frac{\beta_{n,k}}{R_{n,k}}$, it is a non-convex objective.

For the constraints in subproblem $\textbf{SP}_1$, CRB constraint~\eqref{con1} is $\Upsilon_{n,k}^{(1)} \geq \alpha_1$, where $\Upsilon_{n,k}^{(1)} = \frac{\lambda_1 2\pi^2 T^{\text{P}}}{P_{n,k} |\varsigma_k|^2 B_{n,k}}$. The constraint $\Upsilon_{n,k}^{(1)} \geq \alpha_1$ can be transformed into
\[
\frac{\lambda_1 2\pi^2 T^{\text{P}}}{P_{n,k} |\varsigma_k|^2 B_{n,k}} \geq \alpha_1 \implies P_{n,k} B_{n,k} \leq \frac{\lambda_1 2\pi^2 T^{\text{P}}}{\alpha_1 |\varsigma_k|^2} = c.
\]

The set $\{(P_{n,k}, B_{n,k}) \mid P_{n,k} B_{n,k} \leq c\}$ (for some $c>0$) is non-convex. Hence, the feasible region is non-convex.

Therefore, subproblem $\textbf{SP}_1$ has a non-convex objective and non-convex constraints, which is non-convex.

For subproblem $\textbf{SP}_2$, its objective function is given by
\begin{equation}
    \mathcal{E}_{n,k}^{(2)} = I(H_{n,k}, E_{n,k}) \left( \frac{\omega_{n,k}^{(2)} \left( 1 - \frac{1}{1 + \xi_{k} F_{n,k}} \right)}{\frac{\mu F_{n,k}}{C_{n,k}} + \frac{F_{n,k}}{R_{n,k}}} \right).
\end{equation}

Let define $\Gamma_{n,k}' = I(H_{n,k}, E_{n,k}) \omega_{n,k}^{(2)} \left( 1 - \frac{1}{1 + \xi_{k} F_{n,k}} \right)$, Then we can have $
\mathcal{E}_{n,k}^{(2)} = \Gamma_{n,k}' \frac{1}{\frac{\alpha_{n,k}}{C_{n,k}} + \frac{\beta_{n,k}}{R_{n,k}}}$. Since we have proved that $\frac{1}{C_{n,k}}$ is convex and $\frac{1}{R_{n,k}}$ is non-convex, the denominator
$D_{n,k} = \frac{\alpha_{n,k}}{C_{n,k}} + \frac{\beta_{n,k}}{R_{n,k}}$ is non-convex. The reciprocal of a non-convex function is generally non-convex. Therefore, $\mathcal{E}_{n,k}^{(2)}$ is non-convex in the decision variables.

The constraints in subproblem $\textbf{SP}_2$ are analogous to those in subproblem $\textbf{SP}_1$ and thus also non-convex.

Hence, both subproblems $\textbf{SP}_1$ and $\textbf{SP}_2$ are non-convex optimization problems.
\end{IEEEproof}

Based on the above proof, the non-convexity in subproblems $\textbf{SP}_1$ and $\textbf{SP}_2$ primarily arises from three factors. First, the QoE functions contain non-convex terms involving $\frac{1}{R_{n,k}}$. Second, some CRB constraints introduce bilinear terms, e.g., $P_{n,k}B_{n,k}$, making the feasible region non-convex. Third, the objective function in subproblem $\textbf{SP}_2$ includes a fractional form $\frac{\Gamma'_{n,k'}}{\hat{B}_{n,k'}}$, where $\hat{B}_{n,k'}$ depends on $\tfrac{1}{R_{n,k'}}$, further complicating convexity.
 
To handle non-convex term $\tfrac{1}{R_{n,k}}$, we adopt a \ac{sca} framework~\cite{scutari2013decomposition}. At the $t$-th iteration, we have a known solution $(B_{n,k}^{(t)},P_{n,k}^{(t)},C_{n,k}^{(t)})$ and consequently a known $R_{n,k}^{(t)}$. Let consider a function $f(R) = \frac{1}{R}$, where $R=R_{n,k}$. A first-order Taylor expansion around $R_{n,k}^{(t)}$ is derived by
\begin{equation}\label{tay}
    \frac{1}{R} \approx \frac{1}{R_{n,k}^{(t)}} - \frac{R - R_{n,k}^{(t)}}{(R_{n,k}^{(t)})^2}.
\end{equation}
Since $R_{n,k}$ involves $\log_2(1+x)$ with $x=\frac{P_{n,k}|h_{n,k}|^2}{B_{n,k}\sigma_r^2}$, we linearize $\log_2(1+x)$ at $x^{(t)}$, given by
\begin{equation}
\log_2(1+x)=\frac{\ln(1+x)}{\ln(2)}.
\end{equation}
By approximating $\ln(1+x)$ at $x^{(t)}$, we can have
\begin{equation}
\ln(1+x)\approx\ln(1+x^{(t)})+\frac{x-x^{(t)}}{1+x^{(t)}}.
\end{equation}
Therefore, the logarithmic function can be estimated as
\begin{equation}\label{line}
\log_2(1+x)\approx \log_2(1+x^{(t)}) + \frac{x-x^{(t)}}{(1+x^{(t)})\ln(2)}.
\end{equation}

Substituting Eq.~\eqref{line} into $R_{n,k}=B_{n,k}\log_2(1+x)$, we obtain an affine approximation of $R_{n,k}$ around the current iterate, which ensures that the approximation of $\frac{1}{R_{n,k}}$ is also affine in the current iteration. This step is crucial for maintaining convexity at each iteration.
  
For the CRB constraint containing terms such as $P_{n,k}B_{n,k}$, we introduce an auxiliary variable $Y_{n,k}=P_{n,k}B_{n,k}$. To ensure convexity, we apply McCormick envelopes~\cite{mccormick1976computability}, given by
\begin{equation}\left\{
    \begin{aligned}
&Y_{n,k}\geq P_{n,k}\underline{B}+B_{n,k}\underline{P}-\underline{P}\underline{B},\\
&Y_{n,k}\leq P_{n,k}\overline{B},\\
&Y_{n,k}\leq B_{n,k}\overline{P},\\
&Y_{n,k}\geq0,
\end{aligned}\right.\label{crb_bound}
\end{equation}
where $\underline{P},\overline{P},\underline{B},\overline{B}$ are lower and upper bounds derived from the previous iteration. The linear inequalities form a convex polyhedral approximation of bilinear term’s feasible region.
 
In subproblem $\textbf{SP}_2$, we face a fractional objective $\frac{\Gamma'_{n,k'}}{\hat{B}_{n,k'}}$, where $\hat{B}_{n,k'}=\frac{\mu F_{n,k'}}{C_{n,k'}}+\frac{F_{n,k'}}{R_{n,k'}}$. To transform it into a convex function, we introduce an auxiliary variable $\phi_{n,k^{'}}$ and use the quadratic transform, given by
\begin{equation}\label{quad}
    \frac{\Gamma'_{n,k^{'}}}{\hat{B}_{n,k^{'}}}=\max_{\phi_{n,k^{'}}}\{2\phi_{n,k^{'}}\Gamma'_{n,k^{'}}-\phi_{n,k^{'}}^2 \hat{B}_{n,k^{'}}\}.
\end{equation}
Given $\phi_{n,k^{'}}^{(t)}$ from the previous iteration, updating $\phi_{n,k^{'}}$ and linearizing $\hat{B}_{n,k^{'}}$ ensure that the fractional objective is replaced by a form that can be modeled using convex second-order cone constraints. Combined with the preceding linearization steps, the entire problem at each iteration becomes a convex optimization problem.

Under the SCA framework, after applying first-order Taylor expansions to linearize $\frac{1}{R_{n,k}}$ and $\log_2(1+x)$, employing McCormick envelopes for bilinear terms, and using the quadratic transform for fractional objectives, the transformed subproblems $\textbf{SP}_1$ and $\textbf{SP}_2$ at each iteration are convex.

\subsection{The Procedure and Computational Complexity of Proposed Algorithm}
\begin{algorithm}[t]\label{alg2}
\caption{SCA-based Taylor TD (STTD)}
\textbf{Input:} Users' behavioral dynamics, environmental complexity, sensing and communication channel gains, file sizes, and QoE model parameters;

\textbf{Output:} User-level resource allocation decisions $\{B_{n,k}, P_{n,k}, C_{n,k}\}_{k\in\mathcal{K}}$;

\textbf{Initialize:} Taylor TD parameters, replay memory $\mathcal{D}$, policy weight $\bm{\theta}$, target weight $\bm{\theta}^{'}$, initial feasible solution $\{B_{k}^{(0)}, P_{k}^{(0)}, C_{k}^{(0)}, \phi_{k}^{(0)}\}_{k\in\mathcal{K}}$;

\For{each $\text{episode}$ $\mathtt{e} \in \{1,\cdots,\mathtt{E}\}$}
    {
    Reset initial state $\texttt{s}_1$;
    
    \For{\text{each step} $n \in \{1,\cdots,N\}$}
        {
        With probability $\varsigma$ select a random group-level action $\texttt{a}_n$, otherwise, select action based on the Taylor TD policy;

        \For{each user group $i \in \{1, 2\}$}
            {            
            \If{$i == 1$ ($\textbf{SP}_1$)}
                {
                Linearize non-convex components based on Eqs.~\eqref{tay}\eqref{line};

                Conduct bilinear constraints relaxation based on Eq.~\eqref{crb_bound};
                
                Solve the convex subproblem $\textbf{SP}_1$ via SLSQP solver to obtain user-level decision $\{B_{n,k}, P_{n,k}, C_{n,k}\}_{k\in\mathcal{K}_1}$;
                }
            \ElseIf{$i == 2$ ($\textbf{SP}_2$)}
                {
                Make quadratic transformation for fractional objective based on Eq.~\eqref{quad};

                Linearize non-convex components based on Eqs.~\eqref{tay}\eqref{line};

                Solve the convex subproblem $\textbf{SP}_2$ via SLSQP solver to obtain user-level decision $\{B_{n,k}, P_{n,k}, C_{n,k}\}_{k\in\mathcal{K}_2}$;
                }
            }
        
        Observe reward $r_t$ based on average user QoE and transition to new state $\texttt{s}_{n+1}$;

        Store transition $(\texttt{s}_n, \texttt{a}_n, \texttt{r}_n, \texttt{s}_{n+1})$ in replay memory $\mathcal{D}$;

        Prioritize replay to obtain a transition $(\texttt{s}_j, \texttt{a}_j, \texttt{r}_j, \texttt{s}_{j+1})$ from $\mathcal{D}$;

        Calculate the TD target based on Eq.~\eqref{td-1};

        Update action-value function using Taylor correction in Eq.~\eqref{td-2};
        
        Perform the gradient descent operation;

        Every $\Lambda$ steps, update the target network weight with $\bm{\theta}^{'} = \bm{\theta}$;
        }
    }
\end{algorithm}

As shown in Alg.~\ref{alg2}, we present a two-layer data-model-driven algorithm, named \ac{sca}-based Taylor TD (STTD). In the algorithm, group-level and user-level resource allocation decisions are sequentially made. Specifically, in Lines~$5-7$, the Taylor TD neural network makes group-level resource allocation decisions. In Lines~$8-19$, the SCA-based optimization method makes user-level resource allocation for subproblems $\textbf{SP}_1$ and $\textbf{SP}_2$. Then, in Lines~$20-27$, the reward is calculated based on average user QoE, and the Taylor TD neural network is updated based on Taylor correction and gradient descent operations. 

The computational complexity of the STTD algorithm is primarily influenced by the convex optimization for solving subproblems $\textbf{SP}_1$ and $\textbf{SP}_2$. Each convex subproblem involves interior-point methods with a worst-case complexity of \(O(V^3)\), where \(V\) denotes the number of decision variables. This process is repeated for two user groups at each step. Additionally, the Taylor TD updates involve policy network inference and gradient descent, which scale as \(O(L_{\text{policy}})\), where \(L_{\text{policy}}\) represents the operations in the neural network. Replay memory operations, such as prioritized sampling, scale logarithmically with the memory size, which can be expressed as \(O(\log|\mathcal{D}|)\). Overall, the total complexity per episode can be expressed as $O\left(N \cdot \left(2V^3 + L_{\text{policy}} + \log|\mathcal{D}|\right)\right)$. Considering the number of episodes, \(\mathtt{E}\), the total computational complexity of the algorithm is given by $
O\left(\mathtt{E} \cdot N \cdot \left(2V^3 + L_{\text{policy}} + \log|\mathcal{D}|\right)\right)$.

\section{Simulation Results}\label{sim}
In this section, we conduct extensive simulations on the real-world dataset to evaluate the performance of the proposed DA-assisted experience-centric resource management scheme.

\subsection{Simulation Setup}

\begin{figure}[!t]
    \centering
    \includegraphics[width=0.9\mysinglefigwidth]{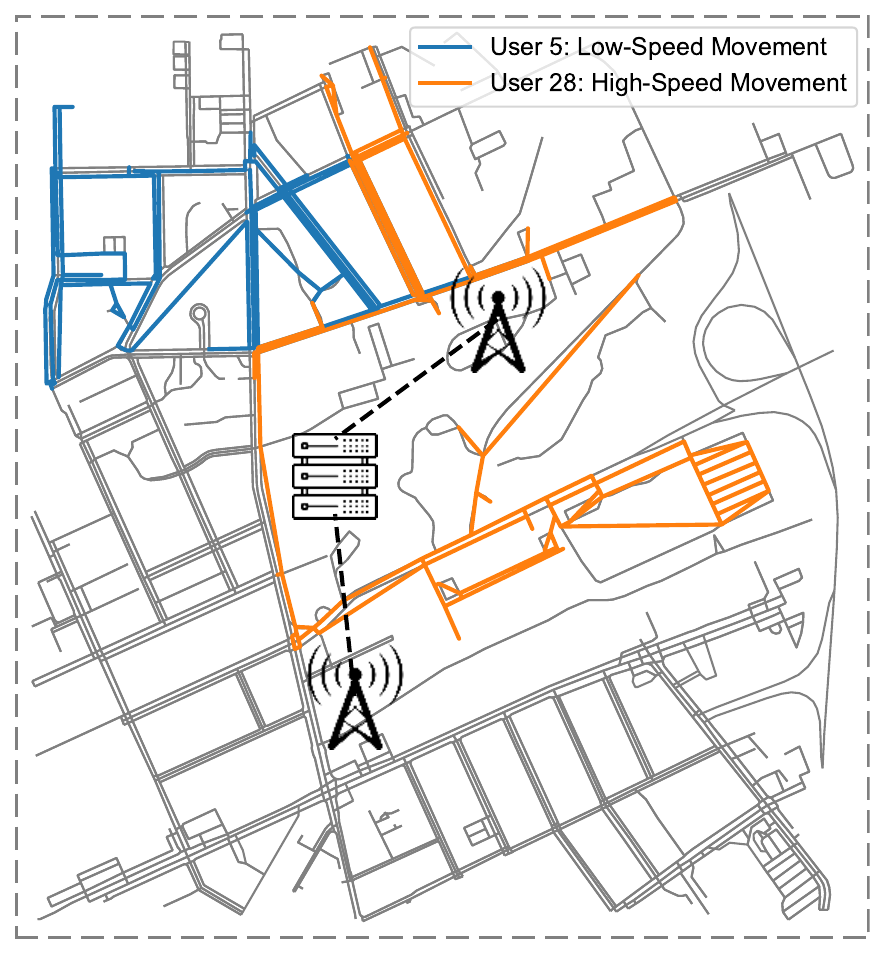}
    \caption{The simulation region in the Waterloo City, ON, Canada.}
    \label{region}
\end{figure}
We consider a simulation region in the Waterloo City, ON, Ontario, as shown in Fig.~\ref{region}. Two base stations equipped with $32$ transmit and receive antennas are deployed at the locations $(43.4661^\circ \mathrm{N}, 80.4789^\circ \mathrm{W})$ and $(43.4601^\circ \mathrm{N}, 80.4820^\circ \mathrm{W})$, respectively. The frequency of carriers of two base stations are $28~\myunit{GHz}$ (n257) and $39~\myunit{GHz}$ (n260) with bandwidth $400~\myunit{MHz}$. The noise power spectral density is set as $-174~\myunit{dBm/Hz}$~\cite{dong2022sensing}. An edge server with computing capacity $15$ G Cycles/s is connected to two base stations. The computing density for processing user tasks is set as $10^7~\myunit{Cycles/MB}$~\cite{huang2024adaptive}. The total number of ISAC users is $30$, where each user moves with different velocities, ranging from $30\sim80~\myunit{Km/h}$. The variance of pathloss for sensing is calculated based on the standard radar equation, i.e., $\hat{\sigma}_{\alpha_k}^2 = {\beta_{\text{RCS}} \lambda_c^2}/{(4\pi)^3 d_k^4}$, where $\beta_{\text{RCS}}$ and $\lambda_c$ are the radar cross section of sensing target and the wavelength of carrier, respectively~\cite{dong2022sensing}. The sensing channel gain of user $k$ is estimated by $\varsigma_k = {\hat{\sigma}_{\alpha_k}^2 \kappa^4 \hat{\epsilon}_k^2 \tilde{\epsilon}_k^2}/{N_\mathtt{r} \sigma_r^2}$~\cite{dong2022sensing, 1597550}. The pathloss for communication is calculated based on $32.4+20\log(d_k)+20\log(f_c)~\myunit{dB}$~\cite{dong2022sensing}. The detailed communication parameter settings are presented in Table~\ref{comm}.

\begin{table}[!t]\caption{Communication Parameters}
\centering
\begin{tabular}{c|c|c}
\hline\hline
\textbf{Parameter} &  \textbf{Description} & \textbf{Value} \\ \hline
$T^\text{Eff}$ & Effective time duration     & $1~\myunit{ms}$~\cite{yan2015simultaneous}                   \\ \hline
$T^\text{P}$   & Pulse width             & $2.5~\myunit{ns}$                                                 \\ \hline
          $B_\text{NN}$ & Null-to-null beam width               & $0.076^\myunit{\circ}$                                                                    \\ \hline
$P$ & Total transmit power & $40~\myunit{W}$ \\ \hline
$\beta_\text{RCS}$ & Radar cross section & 5~$\myunit{m^2}$ \\ \hline
$\hat{\epsilon}$ & Beamforming gain factor of transmit antenna & 1~\cite{dong2022sensing}\\ \hline

$\tilde{\epsilon}$ & Beamforming gain factor of receive antenna & 1~\cite{dong2022sensing}\\ \hline
$\kappa$ & Array gain          & 0.8 \\ \hline
$\alpha_1, \alpha_2, \alpha_3$ & Sensing CRB thresholds & $0.01$                   \\ \hline
$\lambda_1$ & CRB weighting factor of distance &  1 \\ \hline
$\lambda_2$ & CRB weighting factor of velocity & $10^{-20}$                  \\ \hline
$\lambda_3$ & CRB weighting factor of azimuth angle & $10^{-13}$ \\ \hline
\end{tabular}\label{comm}
\end{table}

We consider a short video streaming service in the simulation region, where $10,000$ video sequences are sampled from the YouTube 8M dataset\footnote{YouTube 8M dataset: \url{https://research.google.com/youtube8m/index.html}}. The sampled videos include $8$ categories, i.e., Entertainment, Games, Food, Sports, Science, Dance, Travel, and News. The time duration of each video sequence is $30~\myunit{sec}$, consisting of $15$ equal-length segments. In the \ac{dt} module, user behavioral dynamics mainly consider user swipe behaviors, which are generated based on watching probability distribution\footnote{ACM MM Grand Challenges: \url{https://github.com/AItransCompetition/Short-Video-Streaming-Challenge/tree/main/data}}. The environmental complexity mainly considers user mobility velocity, indicating a high velocity corresponds to a high environmental complexity. The environmental complexity of user $k$ is estimated as $E_k=\exp((v_k^2-v_{\max}^2)/2\rho^2)$ based on the similar principle in~\cite{huangiot}. Parameter $\rho$ equals to $v_{\max}/1.96$, where $v_{\max}$ is set as $80~\myunit{Km/h}$. Users' MOS are generated based on three principles~\cite{shen2024revolutionizing}, latency-based probability distribution $\mathcal{N}(5-0.4\mathcal{R}, 8; 1, 5)$, quality-based one $\mathcal{N}(1+4\mathcal{Q},1;1,5)$, and combined one $\mathcal{N}(1+4\mathcal{Q}-0.4\mathcal{R}, 0.8; 1,5)$, where $\mathcal{R}$ and $\mathcal{Q}$ represent service latency and video quality, respectively. The standard DA data collection frequencies for behavior data, performance data, and environment data are set as $2$, $1$, and $0.5$ per second. The DA data collection attenuation rates of different data attributes are presented in Table~\ref{collection}.

\begin{table}[!t]\caption{STTD Algorithm Parameters}
\centering
\begin{tabular}{c|c||c|c}
\hline\hline
\textbf{Parameter} & \textbf{Value} & \textbf{Parameter} & \textbf{Value} \\ \hline
Hidden layer structure     & $256\times128\times64$ & Batch size & $128$               \\ \hline
Layer connection                & Fully connected                                                  & Initial epsilon & $0.2$ \\ \hline
          Optimizer               & Adam                                                   & Final epsilon & $0.01$                  \\ \hline
Activation function & ReLu & Memory size           & $6,000$ \\ \hline
Maxiter of SLSQP & $150$& Discount factor & $0.9$\\ \hline
Step size of SLSQP & $10^{-4}$ & Learning rate & $10^{-4}$  \\
\hline
\end{tabular}\label{STTD}
\end{table}

To train an effective STTD algorithm, we set $240$ training episodes, each consisting of $60$ steps. In each step, the Taylor TD neural network is responsible for making group-level resource allocation decisions. User-level optimization problems are transformed into convex problems, solved by the SLSQP solver. The reward is then analyzed based on the average user QoE. The Taylor TD neural network is updated based on sampled batches. The detailed STTD algorithm parameters are presented in Table~\ref{STTD}.

We compare the proposed \ac{dt}-assisted resource management scheme with the following benchmark schemes:
\begin{itemize}
    \item \textbf{Without \ac{dt} (w/o \ac{dt})}: A general QoE model consisting of content quality and service quality is utilized. The resource scheduling is based on round robin.
    \item \textbf{Heuristic}: The same \ac{dt}-assisted QoE model as the proposed scheme, but the resource scheduling uses a greedy algorithm.
    \item \textbf{Purely DRL (PDRL)}: The same \ac{dt}-assisted QoE model as the proposed scheme, but the resource scheduling relies on a five-layer \ac{bdqn}~\cite{tavakoli2018action}.
    \item \textbf{Hierarchical Reward-based DRL (HDRL)}~\cite{10136736}: The same \ac{dt}-assisted QoE model, but the resource scheduling relies on a \ac{td3} network, where constraint violations and QoE are used to construct a two-layer reward.
     
\end{itemize}

\subsection{Convergence Performance}
\begin{figure}[!t]
    \centering
    \includegraphics[width=0.8\mysinglefigwidth]{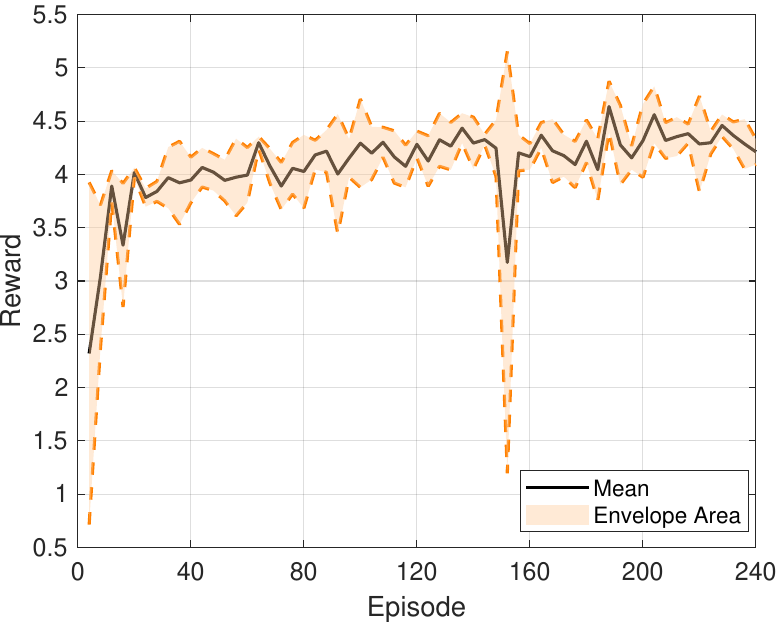}
    \caption{The convergence performance of STTD algorithm.}
    \label{reward_curve}
\end{figure}
As shown in Fig.~\ref{reward_curve}, the convergence performance of the proposed STTD algorithm is evaluated. The envelope area is formulated based on the average reward calculated over every four steps and the corresponding variance. The results show that the reward gradually increases, undergoes minor fluctuations, and ultimately stabilizes around $4.3$. Notably, a temporary decrease in the average reward to $3.2$ is observed at approximately episode $150$, primarily due to extensive explorations. These explorations result in several users' sensing and resource constraints being violated, thereby leading to degraded QoE performance. Overall, the proposed STTD algorithm demonstrates the ability to converge to a high and stable reward, which demonstrates its effectiveness.

\subsection{QoE Probability Distribution}
\begin{figure}[!t]
    \centering
    \includegraphics[width=0.8\mysinglefigwidth]{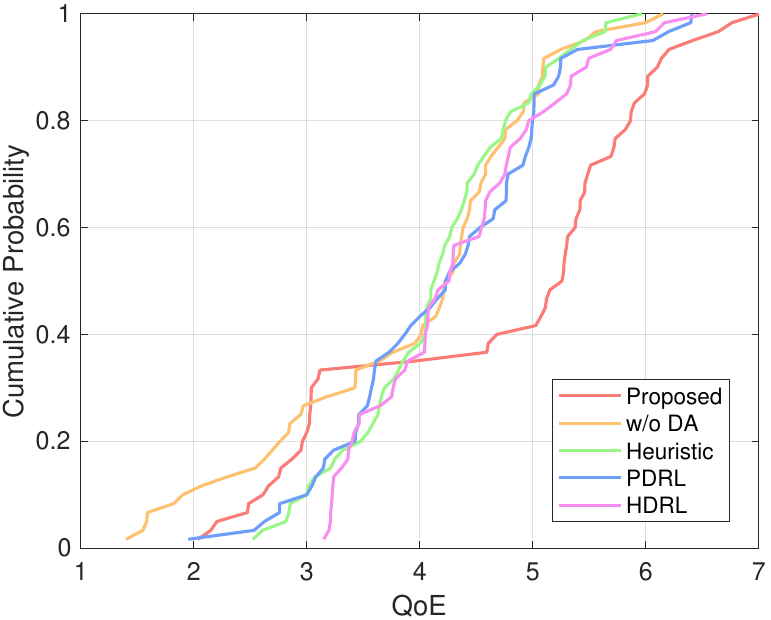}
    \caption{The cumulative probability of QoE.}
    \label{QoE_dis}
\end{figure}
Fig.~\ref{QoE_dis} depicts the cumulative probability distribution of QoE in one episode under different schemes, including the proposed, w/o DA, Heuristic, PDRL, and HDRL schemes. In the early stage, i.e., low QoE range, the proposed scheme does not exhibit superior performance compared to other schemes such as Heuristic and PDRL. This is primarily because the objective of the proposed scheme is to maximize the overall QoE of all users, rather than ensuring fairness across users. Consequently, the proposed scheme sacrifices the QoE performance of some users under poor channel conditions to enable a larger proportion of users to achieve higher QoE. As the QoE increases, this optimization strategy becomes apparent, with the proposed scheme outperforming other methods, as evidenced by its sharper rise and higher cumulative probability at higher QoE values. In contrast, schemes emphasizing fairness, such as HDRL, perform more consistently across all users but fail to achieve the same overall QoE as the proposed scheme. Overall, our proposed scheme can achieve a satisfactory QoE performance compared to other schemes.

\begin{figure*}[!t]
	\centering
	\subfloat[QoE vs. user (400 MHz \& 15 G Cycles/s).]{
		\includegraphics[width=0.32\textwidth]{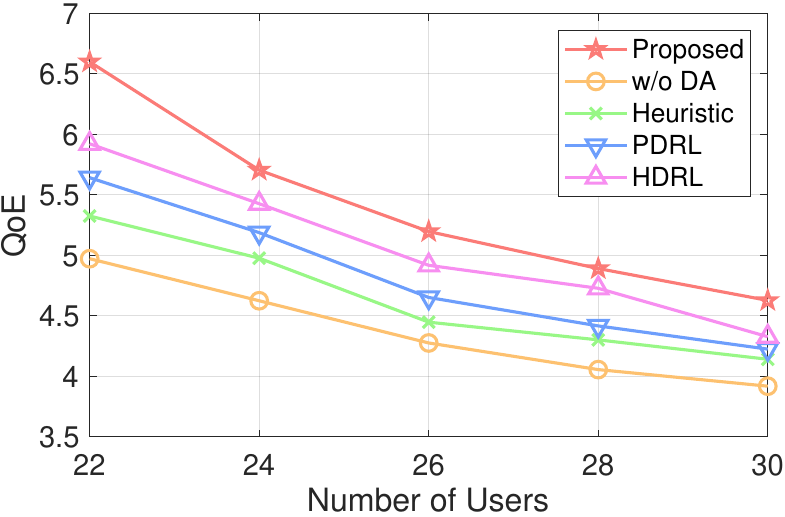}\label{8a}}
	\centering
	\subfloat[QoE vs. bandwidth (30 Users \& 15 G Cycles/s).]{
		\includegraphics[width=0.32\textwidth]{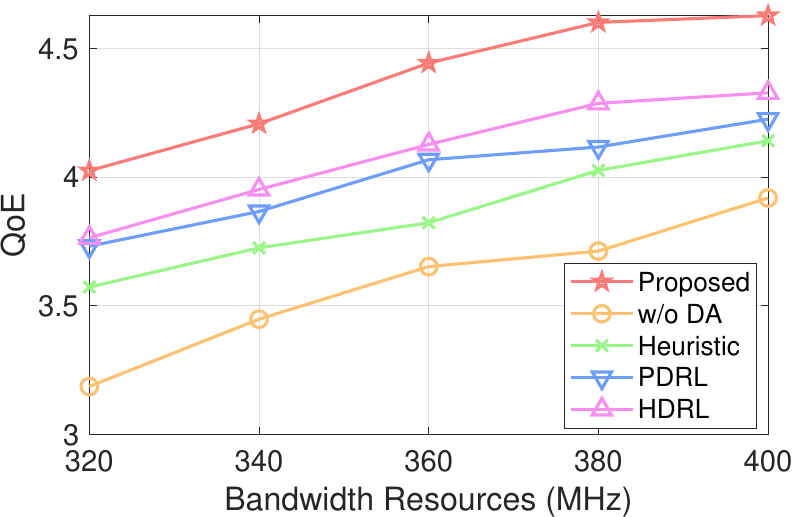}\label{8b}}
	\centering
	\subfloat[QoE vs. computing (30 Users \& 400 MHz).]{
		\includegraphics[width=0.32\textwidth]{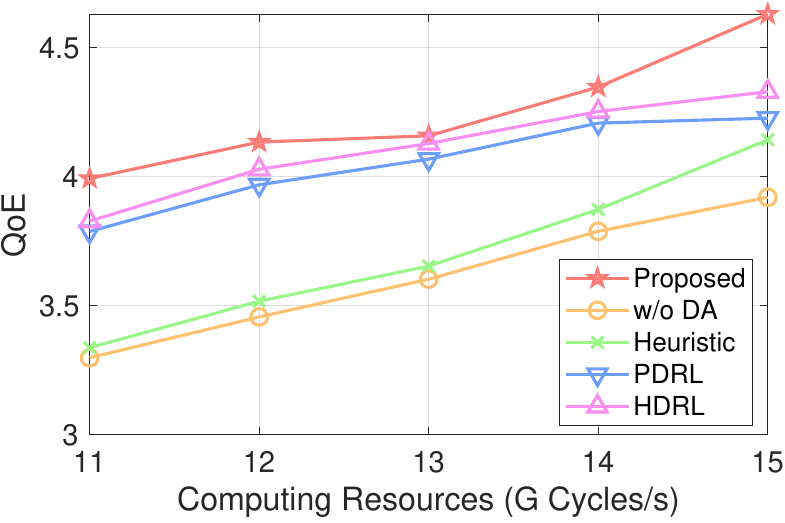}\label{8c}}
  \caption{The QoE performance comparison under different user numbers, bandwidths, and computing resources.}
	\label{fig:components}
\end{figure*}

\subsection{QoE Performance under Different Network Resources}
We also evaluate the QoE performance under varying network resources, as shown in Fig.~\ref{fig:components}. As shown in Fig.~\subref*{8a}, the QoE performance is compared across different numbers of users, with bandwidth and computing resources fixed at $400$ MHz and $15$ G Cycles/s, respectively. It is observed that as the number of users increases, the QoE of all schemes gradually declines. The proposed scheme exhibits a decrease from an initial value of $6.60$ to $4.43$, which degraded QoE up to $32.88\%$. However, it consistently outperforms other schemes, demonstrating the proposed scheme’s ability to adapt effectively to varying user numbers and make appropriate network resource allocation decisions. Furthermore, although the QoE gap between the proposed scheme and other schemes narrows with increasing user numbers, the proposed scheme still achieves a QoE that is $17.86\%$ higher than the w/o DA scheme when the user number is $30$, which further demonstrates the effectiveness of the proposed DA module.

As shown in Fig.~\subref*{8b}, the QoE performance is evaluated under varying bandwidth resources, with the number of users and computing resources fixed at $30$ and $15$ G Cycles/s, respectively. It is observed that as bandwidth resources increase, the QoE of all schemes improves correspondingly. Notably, the QoE performance of the HDRL scheme slightly surpasses that of the PDRL scheme, as the hierarchical reward enables DRL algorithms to explore more efficient network decision-making strategies. However, when the bandwidth resources increase from $380$ to $400$ MHz, the QoE gain of most schemes becomes negligible. This is because the current bandwidth resources are sufficient to provide satisfactory networking services, and further increases in bandwidth do not lead to noticeable improvements in service quality. Consequently, the QoE performance eventually stabilizes at a high value.

\subsection{QoE Performance under Different DA Data Collection Attenuation Rates}

\begin{table*}[!t]
\caption{The QoE performance under different DA data collection attenuation rates.}
\centering
\begin{tabular}{c|cccccc||cccccc||cccccc}
\hline\hline
   \multirow{2}{*}{\textbf{Metric}} & \multicolumn{6}{c||}{\textbf{Attenuation Rate} $\nu_1$}  
   & \multicolumn{6}{c||}{\textbf{Attenuation Rate} $\nu_2$}  
   & \multicolumn{6}{c}{\textbf{Attenuation Rate} $\nu_3$}  \\ 
\cline{2-19} 
    & \multicolumn{1}{c|}{2.00} & \multicolumn{1}{c|}{2.40} & \multicolumn{1}{c|}{2.80} & \multicolumn{1}{c|}{3.20} & \multicolumn{1}{c|}{3.60} & 4.00 
    & \multicolumn{1}{c|}{1.50} & \multicolumn{1}{c|}{1.80} & \multicolumn{1}{c|}{2.10} & \multicolumn{1}{c|}{2.40} & \multicolumn{1}{c|}{2.70} & 3.00 
    & \multicolumn{1}{c|}{0.50} & \multicolumn{1}{c|}{0.70} & \multicolumn{1}{c|}{0.90} & \multicolumn{1}{c|}{1.10} & \multicolumn{1}{c|}{1.30} & 1.50 \\ 
\hline
\textbf{QoE} & \multicolumn{1}{c|}{4.55} & \multicolumn{1}{c|}{4.63} & \multicolumn{1}{c|}{4.57} & \multicolumn{1}{c|}{4.52} & \multicolumn{1}{c|}{4.52} & 4.52 
             & \multicolumn{1}{c|}{4.54} & \multicolumn{1}{c|}{4.59} & \multicolumn{1}{c|}{4.63} & \multicolumn{1}{c|}{4.57} & \multicolumn{1}{c|}{4.53} & 4.50 
             & \multicolumn{1}{c|}{4.48} & \multicolumn{1}{c|}{4.52} & \multicolumn{1}{c|}{4.57} & \multicolumn{1}{c|}{4.63} & \multicolumn{1}{c|}{4.63} & 4.63 \\ 
\hline
\end{tabular}\label{collection}
\end{table*}

We evaluate how DA data collection attenuation rates affect QoE performance under the proposed DA-assisted resource management scheme, as shown in Table~\ref{collection}. Attenuation rates $\nu_1, \nu_2, \nu_3$ represent different DA data collection frequency attenuation rates for behavior data, performance data, and environment data. The number of users, bandwidths, and computing resources are set to $30$, $400$ MHz, and $15$ G Cycles/s, respectively. With the increasing attenuation rate, DA data collection frequency is more sensitive to prediction accuracy, indicating that a low prediction accuracy corresponds to a higher DA data collection frequency and vice versa. It can be observed that when attenuation rates increase, QoE usually increases and gradually increases to a stable state. For instance, when attenuation rate $\nu_1$ increases from $2.00$ to $4.00$, QoE first increases to a top point $4.63$ and then gradually decreases to a stable point $4.52$. The reason for this phenomenon is the trade-off between DA data collection cost and ISAC transmission cost. Specifically, with a higher DA data collection frequency, the DA prediction model can have more realistic data to fine-tune its parameters and achieve high accuracy. The more accurate QoE model can guide more tailored resource management, thus increasing QoE. However, higher DA data collection frequency also occupies more communication resources, which decreases data transmission of ISAC services and thus degrades QoE. The trade-off can usually lead to an optimal attenuation rate to maximize QoE. Additionally, different DA data attributes have various sensitive degrees to attenuation rates. Therefore, selecting appropriate attenuation rates for different DA data attributes to maximize QoE is critical.

\subsection{QoE Performance under Different Sensing Requirements}
As depicted in Fig.~\subref*{8c}, the QoE performance of all schemes is compared under varying computing resources, with the number of users and bandwidth resources fixed at $30$ and $400$ MHz, respectively. It is observed that as computing resources increase, the QoE gap between the proposed scheme and the baseline schemes initially narrows and then widens beyond $13$ G Cycles/s. This phenomenon is primarily attributed to the interplay between computing delay and content quality on the overall QoE. At $13$ G Cycles/s, the sensitivity of QoE to computing resources is lower compared to other points, allowing PDRL and HDRL to achieve similar resource allocation decisions as the proposed scheme. However, when computing resources decrease to $11$ G Cycles/s, the proposed scheme still achieves a QoE that is $4.18\%$ higher than the HDRL scheme. This demonstrates the ability of the proposed STTD algorithm to effectively adapt to varying computing resources and make satisfactory resource allocation decisions.
\begin{figure*}[t]
	\centering
	\subfloat[QoE vs. $\alpha_1$ ($\alpha_2 = \alpha_3 = 0.01$).]{
		\includegraphics[width=0.32\textwidth]{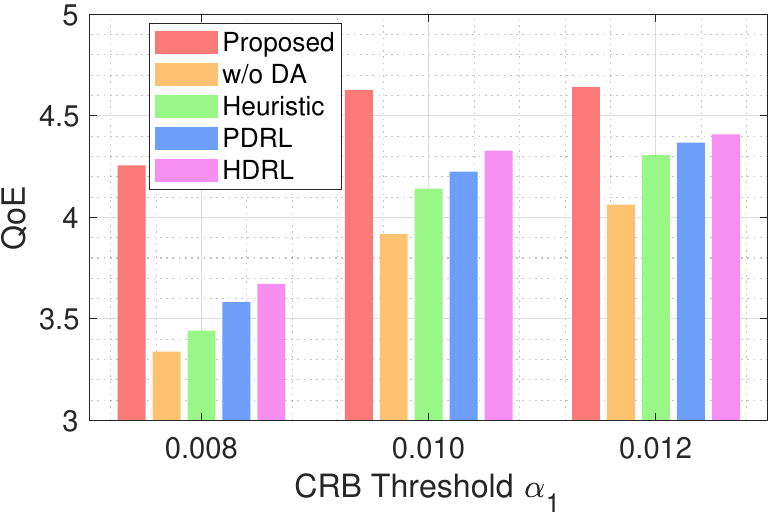}\label{9a}}
	\centering
	\subfloat[QoE vs. $\alpha_2$ ($\alpha_1 = \alpha_3 = 0.01$).]{
		\includegraphics[width=0.32\textwidth]{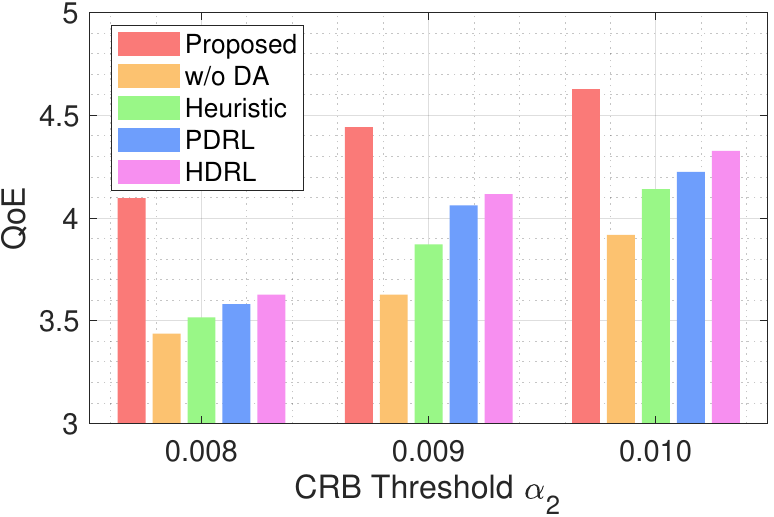}\label{9b}}
	\centering
	\subfloat[QoE vs. $\alpha_3$ ($\alpha_1 = \alpha_2 = 0.01$).]{
		\includegraphics[width=0.32\textwidth]{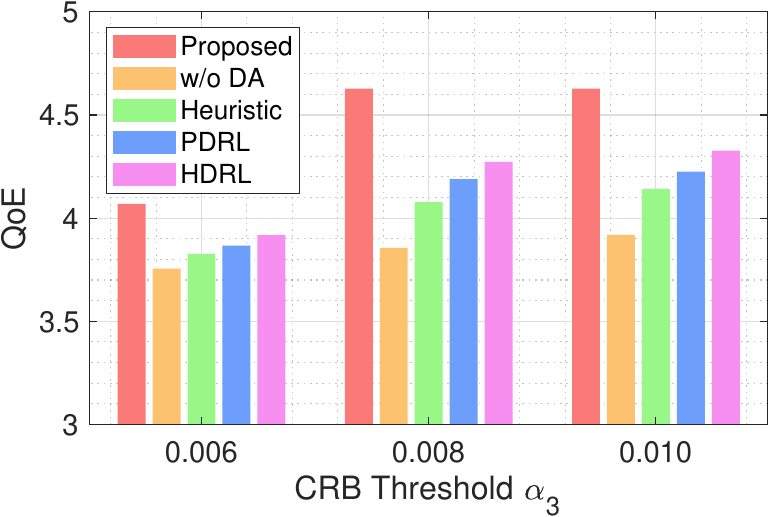}\label{9c}}
  \caption{The QoE performance comparison under different sensing thresholds.}
	\label{fig:QoE_sensing}
\end{figure*}

We further evaluate the QoE performance of different sensing thresholds, as shown in Fig.~\ref{fig:QoE_sensing}. Specifically, Fig.~\subref*{9a} presents the QoE performance comparison of all schemes under different \ac{crb} thresholds $\alpha_1$, with \ac{crb} thresholds $\alpha_2$ and $\alpha_3$ fixed at $0.01$. It can be observed that as the \ac{crb} threshold $\alpha_1$ gradually increases, the QoE also improves and eventually stabilizes. This behavior can be attributed to looser sensing requirements, which reduce communication resource competition, thereby enhancing QoE. Notably, when the \ac{crb} threshold $\alpha_1$ reaches $0.012$, the proposed scheme achieves a QoE that is $14.29\%$ higher than the w/o DA scheme, demonstrating that the proposed DA module effectively enhances user QoE under varying sensing thresholds $\alpha_1$.

As shown in Fig.~\subref*{9b}, we further compare QoE performance under different \ac{crb} thresholds $\alpha_2$, with \ac{crb} thresholds $\alpha_1$ and $\alpha_3$ fixed at $0.01$. It is observed that as the \ac{crb} threshold $\alpha_2$ decreases, the QoE of all schemes gradually declines. The QoE gap between PDRL and HDRL narrows, decreasing from $0.10$ to $0.05$, which indicates that the complexity of the network optimization problem limits both DRL algorithms from achieving superior QoE. The proposed scheme consistently achieves the highest QoE among all schemes. Notably, when the \ac{crb} threshold $\alpha_2$ reaches $0.008$, the QoE of the proposed scheme remains at $3.84\%$ higher than the HDRL scheme, demonstrating the effectiveness of the proposed STTD algorithm under varying sensing requirements.

As depicted in Fig.~\subref*{9c}, the QoE performance of all schemes is further compared under varying \ac{crb} thresholds $\alpha_3$, when \ac{crb} thresholds $\alpha_1$ and $\alpha_2$ are fixed at $0.01$. It is observed that as the \ac{crb} threshold $\alpha_3$ increases, the QoE performance of all schemes gradually converges to a stable value. For instance, the proposed scheme achieves a QoE of $4.63$, while the HDRL scheme stabilizes at $4.33$. Under stringent sensing requirements, e.g., $\alpha_3=0.006$, the QoE of all schemes experiences significant degradation, as additional communication resources are allocated to sensing tasks, reducing the resources available for QoE enhancement. Nonetheless, the proposed scheme consistently outperforms other schemes, achieving QoE values $3.84\%$ and $8.27\%$ higher than those of the HDRL and w/o DA schemes, respectively. These results validate the effectiveness of the proposed STTD algorithm and DA module in handling varying sensing requirements.

\section{Conclusion}
\label{sec:Conclusion}
In this paper, we have proposed a DA-assisted resource management scheme to maximize long-term user QoE in ISAC networks. A DA module has been proposed to manage user-specific QoE models, and the mathematical relationship between resource allocation and sensing accuracy has been modeled via CRB. The proposed DA-assisted resource management scheme can adaptively update user QoE models based on dynamic user behaviors and environments for tailored resource management. For the future work, we will investigate a DA-assisted service-oriented resource management scheme in ISAC networks.

%
\bibliographystyle{IEEEtran}
\bibliography{IEEEabrv,Ref}
\end{document}